\documentclass[twocolumn,showpacs,preprintnumbers,amsmath,amssymb,floatfix]{revtex4}
 \usepackage{graphicx}
 \usepackage{dcolumn}
 \usepackage{bm}

\begin{document}
 \preprint{APS/123-QED}
 \title{Vibration-assisted tunneling through competing molecular states}
 \author{Katja C. Nowack}
 \author{Maarten R. Wegewijs}
 \affiliation{ Institut f\"ur Theoretische Physik - Lehrstuhl A , RWTH
   Aachen , 52056 Aachen , Germany }
 \date{\today}

 \begin{abstract}
   We calculate the non-linear tunneling current through a molecule
   with {\em two} electron-accepting orbitals
   which interact with an {\em intramolecular} vibration.
   We investigate the interplay between Coulomb blockade
   and non-equilibrium vibration-assisted tunneling
   under the following assumptions:
   (i) The Coulomb charging effect restricts the number of
   extra electrons to one.
   (ii) The orbitals are non-degenerate and couple asymmetrically to
   the vibration.
   (iii)  The tunneling induces a non-equilibrium vibrational
   distribution; we compare with the opposite limit of strong
   relaxation of the vibration due to some dissipative environment.
   We find that a non-equilibrium {\em feedback} mechanism in the
   tunneling transitions
   generates strong negative differential conductance (NDC) in the
   model with two competing orbitals, whereas in a one-orbital model it
   leads only to weak NDC.
   In addition, we find another mechanism leading to weak NDC
   over a broader range of applied voltages.
   This pervasive effect is completely robust against
   strong relaxation of the vibrational energy.
   Importantly, the modulation of the electronic transport is based on
   an {\em intra}molecular asymmetry.
   We show that one can infer a non-equilibrium vibrational distribution when
   finding NDC under distinct gate- and bias-voltage conditions.
   In contrast, we demonstrate that any NDC effect in the one-orbital
   case is completely suppressed in the strong relaxation limit.
 \end{abstract}
 \pacs{
   85.65.+h
  ,73.23.Hk
  ,73.63.Kv
  ,63.22.+m
 }
 \maketitle
 Optical spectroscopy of vibrational modes has provided detailed
 information about the structure of molecules~\cite{Herzberg}.
 In a similar fashion, tunneling-current spectroscopy~\cite{Kouwenhoven97rev}
 may reveal microscopic details of single-molecule
 devices~\cite{Park00,Park02,Pasupathy04,Yu04kondo,Yu04ndc,Yu04c60}.
 At low temperature
 electrons tunneling onto a molecular device
 excite {\em discrete} vibrations by spending some of their excess
 energy provided by the bias voltage.
 The resulting changes in the current with bias voltage are determined
 by the effective potentials for the nuclear vibration in different
 {\em charge} states, in contrast to optical spectroscopy.
 This provides information on how the molecule is situated in a
 nanojunction (center of mass modes between the
 electrodes)~\cite{Park02} and the role of different parts of the
 molecule in the transport (internal
 modes)~\cite{Pasupathy04,Yu04c60,Yu04kondo}.
 It is also of interest to {\em control} the electronic
 response of molecular devices by designing their mechanical properties
 through chemical synthesis.
 In this respect, the center of mass modes of a molecule in a
 nanojunction are less attractive since they are difficult to tailor
 and most sensitive to relaxation through interaction with the surface
 of the junction electrodes~\cite{Braig04a,Braig04b}.
 Internal modes of a molecular device, in contrast, 
 are expected to remain more discrete.
 They reflect more intrinsic properties of the device
 amendable to chemical engineering.
 Recently, the coupling of such a tailored mode to the tunneling
 transport has been experimentally investigated in a C$_{140}$
 fullerene-dimer~\cite{Pasupathy04}.
 In view of the ongoing experimental efforts, it is of interest which
 current-spectroscopic features of a molecular device distinguish
 vibrational excitations from electronic ones.
\\
 Theoretical works have already addressed several issues in the weak
 tunneling limit
 by focusing on a one-orbital model including the Coulomb blockade effect
 and coupling to a single localized vibration.
 At finite bias voltage the electron tunneling will
 tend to drive the vibration out of
 equilibrium~\cite{Boese01,McCarthy03, Mitra04b,Koch05a}.
 Several mechanisms may reduce the accumulated vibrational energy:
 coupling to the environment~\cite{Braig04a},
 the tunneling itself~\cite{Mitra04b,Koch04b,Koch05a}
 and also intramolecular vibrational energy redistribution due to
 anharmonic mode coupling.
 The effects of the renormalization of the tunneling by
 Franck-Condon (FC) factors
 on the vibrational distribution were discussed in
 detail~\cite{Mitra04b}.
 These factors incorporate the effect of the change in the nuclear
 configuration when changing the electronic state and charge of the
 molecule.
 In the limit of strong electron-vibration coupling
 current suppression and a related
 super-poissonian current noise were predicted
 as well as a weak NDC effect~\cite{Koch04b,Koch04c}.
 Strong NDC effects~\cite{McCarthy03} were found by assuming
 the electron-vibration coupling to increase with bias voltage.
 Finally, in the strong tunneling limit a Kondo effect due to charge
 (instead of spin) fluctuations was discussed for strong coupling to
 the vibration~\cite{Cornaglia04,Cornaglia05}.
\\
 In this paper we consider a molecule with {\em two}
 electron-accepting orbitals coupled to a single
 internal vibration with frequency $\omega$.
 We consider the  weak  tunneling limit $\Gamma \ll T$
 where $\Gamma$ is the typical tunneling rate to the electrodes and $T$
 the electron temperature.
 The strong Coulomb charging effect is assumed to restrict
 the number of electrons which can be added to the molecule to one.
 This introduces a competition between vibration-assisted tunneling
 processes associated with the different orbitals.
 We focus on the case where the orbitals are non-degenerate (splitting
 $\Delta$) and asymmetrically coupled to the vibration (couplings
 $\lambda_1 \ne \lambda_2$).
 Additionally, we assume that the vibrational distribution on the molecule
 can be driven out of equilibrium (no relaxation).
 For comparison we also discuss the technically simpler limit of
 strong relaxation.
 Due to the asymmetric coupling to the vibration the bare {\em
 electronic} splitting is renormalized to the observable splitting
 $\Delta$ which we will consider as an effective parameter.
 We find that negative differential conductance (NDC) effects are
 enhanced compared with the one-orbital case and actually dominate the
 transport.
 Basically, the orbital coupled stronger to the vibration
 contributes little to the transport but functions as a
 trap due to a {\em feedback} mechanism in the tunneling transitions.
 The competing weakly coupled orbital gives the main contribution to
 the current but at the same time enhances the feedback by providing
 an additional path to the trap.
 Several types of features appear:
 (i) Franck-Condon progressions of alternating conductance resonances and {\em
     anti-resonances} resulting in {\em current oscillations}
 which are robust against strong relaxation;
 (ii) anomalous {\em current peaks} of width $\propto T$ due to a
 strongly bias dependent redistribution of vibrational energy;
 (iii) isolated gate- and  bias- voltage regions
 outside of which the current is strongly suppressed
 due to a stabilization of the charged state which couples stronger
 to the vibration.
 (iv) when the excited orbital couples stronger to the vibration a
 voltage-controlled {\em population inversion} between the two charged
 electronic states takes place which is signalled by NDC occuring at
 special resonances.
 Importantly, the effects are based on the Coulomb blockade and the
 intramolecular asymmetry (non-degenerate electronic states and
  asymmetric coupling to the vibration).
 No asymmetric electronic wave function overlap with the electrodes
 needs to be assumed
 (similar to the NDC induced by to spontaneous emission~\cite{Hettler03}).
 This offers the interesting perspective of designing electronic
 transport properties of a molecular device by synthetic control of the
 electro-mechanical aspects.
 We note that for a related model of two degenerate orbitals current
 rectification was predicted assuming strong
 relaxation~\cite{Flensberg04b}. There the asymmetric coupling to the
 electrodes is essential.
\\
  The paper is organized as follows.
  In Section~\ref{sec:model} we introduce the two-orbital model
  and the master equations for the molecular state occupancies.
  We discuss the qualitative dependence of the FC-factors on the
  vibrational numbers and show that the Condon-parabola from optical
  spectroscopy is also a useful tool in transport-spectroscopy.
  In Section~\ref{sec:single} we first discuss the NDC effect
  noted in Ref.~\cite{Koch04b} for the one-orbital model
  and relate it to the feedback mechanism.
  We then exhaustively discuss the effect of this mechanism for the
  two-orbital case in the limits where one orbital couples weakly to
  the vibration:
  $\lambda_{1}^2 > 1 \gg \lambda_{2}^2$ (Section~\ref{sec:strong_ground}) and
  $\lambda_{1}^2 \ll 1 < \lambda_{2}^2$ (Section~\ref{sec:strong_excited}).
  In Section~\ref{sec:both} we illustrate how the mechanism leads to
  more complex results for asymmetric strong coupling
  $\lambda_{1}^2 \neq \lambda_{2}^2 > 1$.
  We present a comprehensive overview and discussion in
  Section~\ref{sec:discuss}.
\section{Model
  \label{sec:model}
}
\begin{figure}
   \includegraphics[scale=0.325,angle=-90]{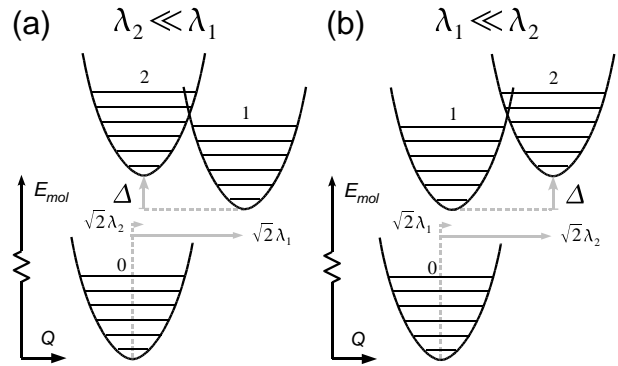}
   \caption{\label{fig:pot}
     Schematic effective nuclear potentials for
     neutral ($i=0$) and charged ($i=1,2$) electronic states of
     molecule.
     (a) Strongly coupled ground state
     (b) Strongly coupled excited state.
   }
 \end{figure}

\begin{figure}
   \includegraphics[scale=0.55,angle=-90]{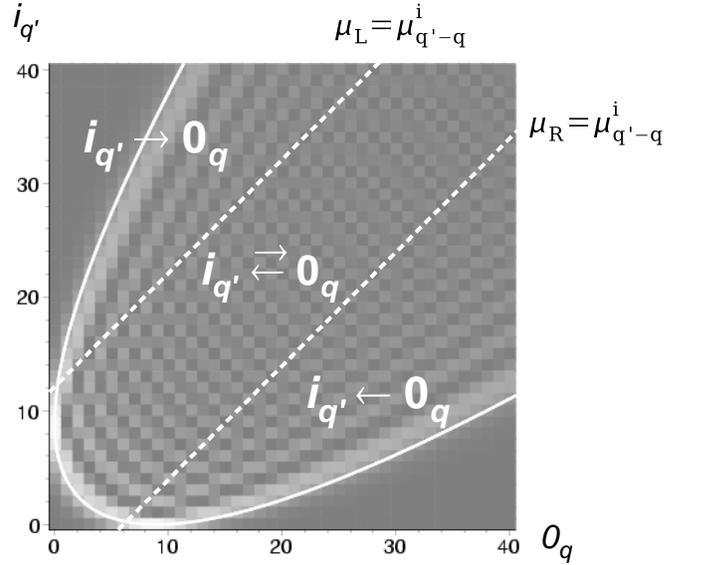}
   \caption{\label{fig:parabola}
     Franck-Condon factors $F^i_{qq'}$ in linear grey-scale
     (white = positive, gray=zero)
     for a transition between $i_{q'}$ and $0_{q}$
     involving a strong relative displacement $\lambda_i =3$.
     The Condon-parabola Eq.~\ref{eq:parabola} (white full line)
     separates the classically forbidden and allowed regions.
     The partial current $I^i_{q'}/\Gamma$, Eq.~(\ref{eq:I_part_i}),
     is obtained by adding the FC-factors for transitions in the
     bias window
     (strip of width and height $V/\omega$ between the dashed white lines)
     horizontally for fixed $q'$.
     The partial current $I^0_q/\Gamma$, Eq.~(\ref{eq:I_part_0}) is
     obtained similarly by adding the FC-factors in this strip
     vertically for fixed $q$
     and adding the contributions from both states $i=1,2$.
   }
 \end{figure}

{\em Molecule.}
 We consider the minimal model $H=H_{M}+\sum_r H_{r}+H_{T}$
 incorporating the molecule ($M$),
 the electrodes $r=L,R$ and the tunneling ($T$) in units $\hbar=k_{B}=1$:
 \begin{eqnarray}
 H_M & = & \sum_{i} \left(
    \epsilon_i n_i + u_i n_{\uparrow}n_{\downarrow}
                   \right)
                   + v n_1 n_2
    \nonumber \\
    &  & + \frac{\omega}{2}\left[
      P^2+\left( Q - \sum_{i} \sqrt{2}\lambda_i n_i \right)^2
     \right]
  ,
  \label{eq:HM}\\
  H_r & = & \sum_{k i \sigma}\epsilon_{k r}a_{k \sigma r}^{\dag}a_{k\sigma r}
  ,
  \label{eq:Hr} \\
  H_T & = & \sum_{k i \sigma r}t_{ir}a_{k \sigma r}^{\dag}c_{i \sigma}+h.c.
  .
  \label{eq:HT}
 \end{eqnarray}
 The Hamiltonian $H_M$ describes a molecule with two
 electron-accepting orbitals $i=1,2$ (operators $c_{i\sigma}$, energies
 $\epsilon_i$) with an energy splitting $\Delta = \epsilon_2 -\epsilon_1$.
 Here $n_{i}= \sum_{\sigma} n_{i\sigma}$
    , $n_{i\sigma}= c_{i \sigma}^{\dag}c_{i \sigma}$.
 We assume throughout that double occupation of each orbital is
 completely suppressed due to strong local Coulomb interactions $u_i$ i.e.
 $n_{i} \leq 1$.
 Simultaneous occupation of the two orbitals is similarly suppressed
 by the interaction $v$ which introduces an important correlation:
 $\sum_i n_i \leq 1$.
 Since $u_i$ and $v$ are the largest energy scales they do not enter
 explicitly in any further way.
 The nuclear configuration of the molecule is assumed to be altered
 with respect to some coordinate $Q$ when either orbital is occupied.
 The effective potentials for the nuclear motion in these
 charged states (here in the harmonic approximation, frequency
 $\omega$) are shifted relative to the neutral state by
 $\sqrt{2}\lambda_{i},i=1,2$.
 The coordinate $Q=(b+b^{\dagger})/\sqrt{2}$ is normalized to the
 nuclear zero-point motion by $(M \omega)^{-1/2}$ ($M=$ nuclear mass
 involved) and $\lambda_i$ is dimensionless.
 Here $b^{\dag}$ excites the vibrational mode by one quantum $\omega$
 and $P=(b-b^{\dagger})/\sqrt{2}i$ is the conjugate nuclear momentum.
 The energy scale characterizing the electron-vibration coupling
 associated with orbital~$i=1,2$ is the change in the elastic
 energy at fixed nuclear configuration $\omega \lambda_i^2$.
 Vibration-assisted processes are thus expected to lead to a progression
 of conductance resonances spread over a bias voltage range of at least
 $\sim \max \{\lambda_i^2 \omega \}$.
 Typical energies $\omega$ of internal vibrations observed experimentally
 range up to a few tens of meV~\cite{Park02,Pasupathy04,Yu04kondo,Yu04ndc,Yu04c60}.
 In general electron-vibration coupling is expected to be particularly
 strong for many-particle states of molecules which are
 characterized to a good approximation by occupation of a particular
 orbital~\cite{Koeppel84}, which is typically the case for 
 charged states of otherwise neutral molecules.
 Large relative displacements $|\lambda_i| > 1$ of the nuclear potentials
 may be expected, for instance,
 when $Q$ is an angle coordinate and the nuclear configuration
 of the charged molecule is internally twisted relative to the neutral
 one~\cite{Cizek04}. 
\\
 The electronic energy parameters in Eq.~\ref{eq:HM}
 are the relevant effective parameters for finite $\lambda_i$.
 These are related to their values for
 $\lambda_1=\lambda_2=0$ (indicated by a superscript (0)) by
 $\epsilon_i = \epsilon_i^{(0)} - \lambda_i^2         \omega,
  u_i        = u_i^{(0)}       -2 \lambda_i^2         \omega,
  v          = v^{(0)}         -2 \lambda_1 \lambda_2 \omega$.
 This is seen by diagonalizing the molecular Hamiltonian through a
 translation of the nuclear coordinates~\cite{LangFirsov63},
 $U=\prod_{i=1,2} e^{-\lambda_i n_i (b^{\dag}-b)}$.
 The resulting Hamiltonian has the form~(\ref{eq:HM}) where
 $\lambda_i,i=1,2$ is eliminated.
 The electron becomes ``dressed'' with vibrational excitations
 (polaron) resulting in the renormalization of
 the energy parameters which we anticipated in writing Eq.~\ref{eq:HM}.
 The renormalization of the charging energies is irrelevant here since we
 assume them to be largest energy scales (i.e.
 $ u_i^{(0)} \gg      2 | \lambda_i |^2       \omega,i=1,2$
 and
 $ v^{(0)}   \gg      2 |\lambda_1 \lambda_2 |\omega$).
 The correlations $n_1+n_2 \leq 1,n_i \leq 1$ are thus not weakened.
 Cases where a strong renormalization of the interaction becomes
 relevant were discussed in~\cite{Cornaglia04,Cornaglia05}.
 In contrast to the one-orbital model the renormalization of 
 the orbital energies is important when the coupling to the vibration
 is asymmetric, $\lambda_1 \neq  \lambda_2$.
 Then the electronic splitting is renormalized to an effective value
 $\Delta =\epsilon_2-\epsilon_1 =
 \Delta^{(0)}+\omega(\lambda_1^2-\lambda_2^2)$
 which can even have a different sign as $\Delta^{(0)}$.
 Since only the excitation energy $\Delta$ is observable in the
 transport characteristics, we use it as an independent positive
 parameter i.e. the state $1$ by definition has the lowest
 renormalized energy $\epsilon_{1}$.
 We are interested in the case where resonances related to orbital and
 vibrational excitations occur on the same voltage scale, i.e.
 $\omega \sim \Delta \lesssim \max \{\lambda_i^2 \omega \}$.
 The transport mechanism which we wish to illustrate operates
 in the limit of asymmetric coupling.
 This requires that either the lowest orbital couples strongly to the vibration
 or the excited orbital, see Fig.~\ref{fig:pot}.
 We point out that in Hamiltonian (\ref{eq:HM}) we have not written
 intramolecular terms which couple the two nuclear potentials 1 and 2.
 Such terms become important, for instance, when the electronic energy
 splitting is an integer multiple $p$ of the vibrational energy
 quantum, $\Delta = p \omega$.
 This has been discussed in Ref.~\cite{Flensberg04b} for the
 case $\Delta=0$ and $\lambda_1=-\lambda_2$.
 Here we avoid such degeneracies, i.e. we assume $t \ll min_p
 \{ \Delta-p\omega \}$, where $t$ is a tunneling amplitude between the
 electronic states.
 Furthermore, we can safely disregard electronic transitions
 induced by the nuclear motion near the crossing of the potentials 1
 and 2 since for large asymmetric coupling the barrier separating the
 minima of potentials $i=1,2$ is
 $(\Delta/\omega (\lambda_1-\lambda_2)^{-1} \pm
                 (\lambda_1-\lambda_2)          )^2 \omega/4 \gg \omega
 $.
 Below we will also present results for cases of moderate asymmetry of the
 vibrational couplings where such effects may start to play a
 role. These will serve as a simple starting point for the discussion
 of the strong asymmetry case and also illustrates the enhancement of NDC
 effects when multiple orbitals (instead of just one) are competing in
 the transport.
\\
 The electrodes $r=L,R$ are modeled by $H_r$, Eq.~(\ref{eq:Hr}), as
 non-interacting quasi-particle reservoirs at electro-chemical
 potential $\mu_r$.  The electrode-molecule tunneling $H_T$,
 Eq.~(\ref{eq:HT}), picks up the shift of the nuclear
 coordinate from the unitary transformation of the molecular operators:
 $H_T=
 \sum_{k i \sigma r}
   t_{ir}a_{k\sigma r}^{\dag} e^{-\lambda_i (b^{\dag}-b)}
   \bar{c}_{i\sigma}+h.c.$
 Here $\bar{c}^{\dag}_{i \sigma}$ creates a polaron state
 associated with the effective potential of electronic orbital~$i=1,2$.
 Since we consider here an intramolecular vibration we do not include
 a dependence of the bare tunneling matrix elements $t_{ir}$ on the
 coordinate $Q$ (shuttle-effect, cf. Ref.~\cite{McCarthy03,Koch04a}).
\\
 {\em Master equations.} We are interested in the weak tunneling
 regime, $\Gamma \ll T$,
 where in addition the vibrational excitations can be resolved,
 $T \ll \omega$.
 We can describe the transport using diagonal density matrix elements
 $P^i_{q}$ (occupation probabilities, $\sum_{i=0}^2 \sum_{q} P^i_q=1$) ,
 where $q=0,1,2,\ldots$ is the vibrational number
 and $i=1,2$ denotes the charged state with only orbital~$i$
 occupied and the neutral state is labeled by $i=0$.
 The transitions including the vibrational
 excitations $q,q'$ are denoted by $0_q
 \leftrightarrows 1_{q'}$ and $0_q \leftrightarrows 2_{q'}$.
 The occupation probabilities are coupled by the stationary master
 equations
 \begin{eqnarray}
 & \dot{P}_q^0   =0&=\sum_{i}\sum_{rq'}\left(
       W_{0q\leftarrow iq'}^{r}P^i_{q'}-2W_{iq'\leftarrow 0q}^{r}P^0_q
      \right)
     ,
     \nonumber \\
 & \dot{P}_q^{i} =0&=        \sum_{r q'}\left(
     2W_{iq\leftarrow 0q'}^{r}P^0_{q'}- W_{0q'\leftarrow iq}^{r}P^i_q
      \right)
    ,
\label{eq:dotP}
 \end{eqnarray}
 where $i=1,2$, with transition rates ($f_r(\epsilon) \equiv
 (e^{(\epsilon-\mu_r)/T}+1)^{-1}
$)
 \begin{eqnarray}
 W_{iq' \leftarrow 0q}^{r} & =& \Gamma^{ri}_{qq'}    f_{r}( \mu^{i}_{q'-q} )
 ,
 \nonumber \\
 W_{0q \leftarrow iq'}^{r} &= & \Gamma^{ri}_{qq'}( 1-f_{r}( \mu^{i}_{q'-q} ) )
 .
 \label{eq:rate}
 \end{eqnarray}
 The addition energies for the transition $i_{q'}\leftarrow 0_q$ are
 \begin{equation}
   \mu^{i}_{q'-q}=\epsilon_i+(q'-q)\omega -\alpha V_g
   .
 \end{equation}
 The gate voltage $V_g$ effectively varies $\mu$ relative to
 the ground-state transition energy $\mu^{1}_{0}$ ($\alpha =$
 capacitance ratio) and
 the bias voltage $V>0$ is applied symmetrically,
 $\mu_{L,R}=\mu \pm V/2$.
 The stationary current flowing out of reservoir
 $r=L,R$ is given by
 ($I_L+I_R=0$)
 \begin{equation}
   I_r=\sum_{q q'}\sum_{i} \left(
     2W_{iq\leftarrow 0q'}^{r}P_{q'}^0
     -W_{0q\leftarrow iq'}^{r}P_{q'}^i
   \right) .
   \label{eq:I}
 \end{equation}
 The equations for the one-orbital case with coupling $\lambda_1$ are
 obtained by simply discarding all $P^2_{q'}$ in Eqs.~(\ref{eq:dotP})
 and~(\ref{eq:I}) and are equivalent to those in
 Refs.~\cite{Boese01,McCarthy03,Mitra04b,Koch04b,Koch05a}.
 The current will change whenever a line in the $(\mu,V)$ plane is
 crossed corresponding to a right-electrode resonance
 $\mu_R=\mu^{i}_{q'-q}$ (positive slope in $(\mu,V)$ plane)
 or a left-electrode resonance
 $\mu_L=\mu^{i}_{q'-q}$ (negative slope).
 Importantly, due to the harmonic excitation spectrum
 only the {\em change} in vibrational number $q'-q$ enters in the
 resonance condition:
 transitions between all states $i_{q'}$ and $0_q$
 with fixed difference $q'-q$ become allowed at a single resonance.
 A {\em cascade} of single-electron transitions
 can then lead to a significant population of high vibrational
 excitations, e.g.
 $i_0 \rightarrow  0_0 \rightarrow
  i_1 \rightarrow  0_1 \rightarrow
  i_2 \rightarrow  0_2 \rightarrow \cdots$
 is a possible cascade for
 $\mu_L > \mu^i_1 > \mu^i_0 > \mu_R$.
\\
 {\em Franck-Condon factors and Condon parabola.}
 The tunneling rates consist of two factors:
 $\Gamma^{ri}_{qq'} =\Gamma^{ri} F^i_{qq'}$.
 The tunneling rates $\Gamma^{ri} = 2\pi |t_{ir}|^2 \rho_r$ between
 electrode $r=L,R$ (density of states $\rho_r$) and orbital~$i=1,2$
 determine the overall current scale.
 The FC factors $F^{i}_{q'q}=F^{i}_{qq'}$ take into account that the
 stable nuclear geometry is changed when occupying orbital~$i$:
 \begin{eqnarray}
   F^i_{q q'}
  = |\langle q |X_i|q' \rangle|^2
  = e^{-\lambda_i^2} \frac{q!}{{q'}!} \lambda_i^{2|q-q'|}
   \left( L^{|q-q'|}_{q}(\lambda_i^2 )\right)^2
  ,
 \end{eqnarray}
 where $L$ is the associated Laguerre-polynomial and $q<q'$.
 Note that the sign of $\lambda_i$ does not play a role.
 The general sum rule
 $\sum_q F^{i}_{qq'}=\sum_{q'} F^{i}_{qq'}=1$,
 guarantees that the current will saturate at large bias voltage to
 the value it would have without the vibrations (``electronic
 limit'').
 This holds {\em only} when the $\lambda_i$ do not depend strongly on
 the bias voltage (cf.~\cite{McCarthy03}).
\\
 Without vibrations, asymmetry of the tunneling rates with respect to
 the orbital- and electrode- index gives rise to
 NDC~\cite{Hettler02nanobio} and super-poissonian current
 noise~\cite{Thielmann04b},
 see also~\cite{cottet04prl,cottet04prb,belzig05prb}.
 Below we show that 
 qualitatively different dependence of the FC factors $F^i_{qq'}$ on
 the vibrational numbers for state $i=1,2$
 and the effective energy splitting $\Delta$ 
 give rise to NDC effects which can dominate the transport.
 We therefore set $\Gamma^{ir}=\Gamma,i=1,2,r=L,R$
 and restrict our discussion to $V>0$ since in this case $I(-V)=-I(V)$.
 We symmetrize the stationary current $I=(I_L-I_R)/2$
 and decompose it into a sum of positive partial currents of the states
 weighted with their occupation:
 \begin{eqnarray}
   I &=&
            \sum_{q}  I^0_q P^0_q    +
  \frac{1}{2}\sum_{i} \sum_{q'} I^i_{q'} P^i_{q'}
   \label{eq:I_sym} \\
   I^0_q &=& \Gamma
   \sum_{q'} \sum_{i}
    F^i_{q q'}
    \left( f_{L}(\mu^{i}_{q'-q}) - f_{R}(\mu^{i}_{q'-q}) \right)
   \label{eq:I_part_0} \\
   I^i_{q'} &=& \Gamma
   \sum_q
    F^i_{q q'}
    \left( f_{L}(\mu^{i}_{q'-q}) - f_{R}(\mu^{i}_{q'-q}) \right)
   \label{eq:I_part_i}
 \end{eqnarray}
 For low $T \ll \omega$ the partial currents are  the FC
 factors $F^i_{qq'}$ summed over the transitions $0_q \leftrightarrows
 i_{q'}$ inside the bias window $\mu_L > \mu^i_{q'-q} > \mu_R$
 in Fig.~\ref{fig:parabola}.
 Note that the partial current of the neutral state $I^0_q$
 has contributions from both charged states $i=1,2$ into which it can
 decay.
 One can understand the numerical results in almost all
 detail using the following simple graphical scheme.
 This approach works for multiple orbitals and also for a more general
 shape of the nuclear potential.
 Without going into the details, we comment on the basic points in the
 procedure.
 In Fig.~\ref{fig:parabola} the FC factor associated with the transition
 between a pair of states $0_q \leftrightarrows i_{q'}$
 is depicted  as function of $q$ and $q'$.
 The change in the partial currents in (\ref{eq:I_sym}) with
 increasing bias can be understood by drawing the bias
 voltage window in this figure.
 Only qualitative features of the FC factors  are of importance
 which follow from simple quasi-classical arguments.
 The FC-factor $F^i_{q q'}$ is basically non-zero only in the
 classically allowed region delimited by the tilted {\em Condon
   parabola}~\cite{Herzberg}:
 \begin{equation}
   \label{eq:parabola}
    q+q' \geq  \frac{|q-q'|^2}{\lambda_i^2} +\frac{\lambda_i^2}{2}
 \end{equation}
 In this region the classical orbits of the nuclear motion in the
 shifted potentials intersect in phase-space.
 The FC-factor oscillates with $q,q'$ with a
 quasi-classical envelope which varies algebraically on the scale of
 $\lambda_i^2$~\cite{Perelomov}.
 The global maxima $\approx 1/\lambda_i^2$ are attained where the
 parabola touches the axes ( $q \approx \lambda^2_i,q'=0$ and $q=0,q'
 \approx \lambda^2_i$).
 In the classically forbidden region where the opposite
 of condition~(\ref{eq:parabola}) holds the FC factor $F^i_{qq'}$ is
 exponentially small.
 In the forbidden regions where $q' \gg q$ or $q' \ll q$ the nuclear
 momenta of the two motions are incompatible.
 These regions exist for any value of $\lambda_i$.
 Starting from the allowed region the FC-factors eventually decrease
 exponentially with increasing $q'$ or $q$.
 On the other hand, the forbidden region where $q,q' \ll \lambda_i^2$
 is only well defined for $\lambda_i^2 \gg 1$.
 In this case $F^i_{qq'}$ initially {\em increases exponentially} with
 increasing $q'$ or $q \ll \lambda_i^2$.
 This ``inverted'' regime
 exists for nuclear potentials of more general shape
 with large relative displacements of the potential minima.
\\
 As the bias window in Fig.~\ref{fig:parabola} widens, the partial
 currents increase.
 For weak coupling $\lambda_i^2 \ll 1$ the Condon-parabola is very
 narrow i.e. transitions which conserve the vibrational number have the
 largest amplitude.
 When the bias window reaches the vertex of this narrow parabola
 nearly all  partial currents reach their maximal value at once.
 For strong coupling, $\lambda_i^2 \gg 1$, the
 parabola is very broad and the partial currents show a slow exponential
 increase as the bias window widens.
 The complex transport characteristics of the two orbital model
 considered here follow from two intramolecular asymmetries between
 the orbitals:
 (1) the bias window covers different parts of the Condon-parabola due to the
 electronic splitting $\Delta$ and
 (2) the Condon-parabolas are qualitatively different due to
 asymmetric coupling to the vibration.
 \\
{\em Strong relaxation.}
 We now prove an important restriction on the occurrence of NDC in the
 limit where the vibrational excitations completely relax before
 each tunneling event due to some dissipative environment. This limit implies
 the factorization ansatz $P_q^i=P^i P_q$ with the vibrational
 equilibrium distribution
 $P_q=e^{-q\omega/T}(1-e^{-\omega/T})$.
 We can then reduce the equations (\ref{eq:dotP}) to an effective
 electronic three-level problem with effective bias-voltage dependent
 rates (cf.~\cite{Braig04b,Mitra04b}) obtained by
 averaging over the equilibrium distribution:
 \begin{equation}
   \label{eq:rate_eq}
   W^r_{0 \leftarrow i}=\sum_{qq'} W^r_{0q\leftarrow iq'} P_{q'},
   W^r_{i \leftarrow 0}=\sum_{qq'} W^r_{iq\leftarrow 0q'} P_{q'}
 \end{equation}
 (these vary monotonically with $V$) and
  $W_{0 \leftarrow i} = \sum_r W^r_{0 \leftarrow i}$
 ,$W_{i \leftarrow 0} = \sum_r W^r_{i \leftarrow 0}$.
 The stationary current in electrode $r=L,R$ reads
 ( $I_L+I_R=0$):
 \begin{eqnarray}
   I_r= 2
   \frac{
     \sum_{i}
     \left(
       W_{i\leftarrow 0}^{r}
     - W_{0\leftarrow i}^{r}
            \frac{W_{i \leftarrow 0}}
                 {W_{0 \leftarrow i}}
     \right)
   }
   {
     1+2 \sum_{i}
             \frac{W_{i \leftarrow 0}}
                  {W_{0 \leftarrow i}}
   }
  \label{eq:Ieq}
  .
 \end{eqnarray}
 In Appendix~\ref{app:relax} we show that in this case for two
 orbitals the current can be reduced by increasing the bias voltage at
 resonances related to the
 {\em left electrode}, $\mu_{L}=\mu^i_{k},k=1,\ldots$ (for $V>0$),
 i.e. lines with positive slope in the $(\mu,V)$ plane.
 Here the transition $i_{q+k} \leftarrow 0_q$ becomes allowed.
 Any NDC along a resonance line with positive slope is thus a 
 {\em proof of a non-equilibrium vibrational distribution} on the
 molecule.
 For one orbital the current can never decrease with $V$ in this limit.
 Finally, we consider the limit where in addition to the strong
 relaxation, the transitions $0_q \leftrightarrows i_{q'}$,
 $i=1,2$ are not correlated i.e. the renormalized Coulomb
 interaction~\cite{Cornaglia04,Cornaglia05} is zero: $v'=0$.
 It is readily shown (Appendix~\ref{app:uncorr}) that in this case the
 current increases monotonically with bias voltage for {\em any number} of
 orbitals. The strong Coulomb correlations are thus essential for
 NDC effects.
\\
{\em Intermediate relaxation.}
 Basically all the physics is captured by considering the opposite limits of
 negligible and strong relaxation. We have confirmed this by
 considering intermediate regimes where we add a relaxation term
 $\sum_{q'} W_{q \leftarrow {q'}} P^i_{q'}$ 
 to the right-hand side of the equation for $\dot{P^i_q}$, Eq.~(\ref{eq:dotP}).
 We considered an environment~\cite{Boese01} with either ohmic
 ($s=1$) or sub-ohmic ($s=0$) spectral function $J(E)=\gamma
 |E/\omega|^s$ for which the relaxation rates are
 $W_{q \leftarrow {q'}} =J(\omega (q-q')) [ \pm N(\omega(q-q')) ]$
 for $q \gtrless q'$  where $N(E)=(e^{\beta E}-1)^{-1}$.
 We briefly discuss the results for intermediate relaxation when interesting
 deviations from a simple interpolation between non-equilibrium and
 strong relaxation limit occur.
\section{Results
\label{sec:results}}
 We now present results for the stationary current $I$
 (Eq.~(\ref{eq:I})) and differential conductance $dI/dV$ for
 symmetric tunneling rates $\Gamma^{ir}=\Gamma,r=L,R,i=1,2$.
 Throughout the paper we set the temperature to $T=0.025 \omega$.
 Gray-scale plots of $dI/dV(\mu,V)$ have been given different linear
 scale factors for $dI/dV \gtrless 0$ to clarify the voltage
 conditions under which NDC occurs.
 The NDC magnitude can be inferred from the presented $I(V)$ curves.
 We note that the results for a molecule with two neutral states and
 one charged state are simply obtained by inverting the sign of
 $\mu-\mu^1_0$ and modifying the discussion accordingly.
\subsection{Strongly coupled single orbital
- Feedback mechanism and  weak NDC
\label{sec:single}
}
\begin{figure}
  \includegraphics[scale=0.7]{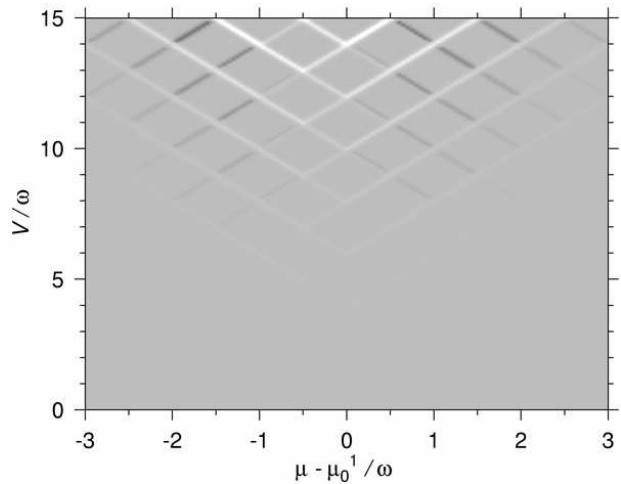}
   \caption{\label{fig:G_one_L}
     One-orbital model with strong electron-vibration coupling $\lambda_1=5.0$.
     Differential conductance in gray-scale (gray: $dI/dV=0$, white /
     black: $dI/dV \gtrless 0$) as function of bias $V$ and gate
     voltage $\mu$.
   }
 \end{figure}

 In Ref.~\cite{Koch04b} the one-orbital model in the limit of strong
 coupling to the vibration ($\lambda_1^2 \gg 1$) was shown to exhibit
 a current suppression 
 which was related to ``avalanches'' of vibration
 assisted tunneling processes which also leads to super-poissonian
current noise effects~\cite{Koch04b,Koch04c}.
 Additionally, for asymmetric gate voltages 
 a weak NDC effect was noted in the absence of relaxation.
 Here we focus on this weak NDC effect and additional small current {\em peaks} (not
 discussed in Ref.~\cite{Koch04b}) which are
 visible in Fig.~\ref{fig:G_one_L} as black lines and white-black
 double lines, respectively.
 The mechanism responsible for this is based on an energy asymmetry
 (induced by the gate potential) and the qualitative features of the
 FC-factors.
 This mechanism will also play a role in the transport
 involving two (or more) orbitals where it results in much stronger
 effects.
 We therefore consider this simple case in some detail.
\\
 \begin{figure}
  \includegraphics[scale=0.35,angle=0]{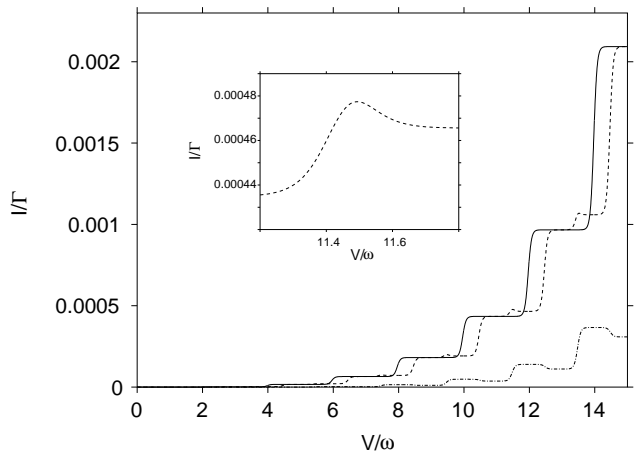}
  \caption{\label{fig:I_one_L}
    Current-voltage characteristics for
    increasing $\mu-\mu^1_0= 0   \omega,
                             0.25 \omega,
                             1.75 \omega$ (going down)
    in Fig~\ref{fig:G_one_L} for non-equilibrium vibrations
    (Eq.~(\ref{eq:I})).
    Inset: small current peak for $\mu-\mu^1_0= 0.25\omega$ which for
    larger $\mu-\mu^1_0$ develops into a current drop.
  }
\end{figure}
{\em Current suppression.}
 The differential conductance plotted in Fig.~\ref{fig:G_one_L} is
 symmetric about $\mu=\mu^1_0$ up to unimportant differences in
 amplitude due to spin degeneracy of the orbital (without which we
 would have exact symmetry) so
 we only discuss the case $\mu \geq \mu^1_0$; the opposite case
 follows from interchanging the roles of the electronic states $0
 \leftrightarrow 1$.
 For low voltages $V \ll 2 \omega \lambda_1^2$ states with low
 vibrational number are predominantly occupied and the current is
 exponentially suppressed both in the limit of weak and strong
 relaxation~\cite{Koch04b,Koch04c}.
 This is related to the classically forbidden
 region $q,q' \ll \lambda_1^2$ in Fig.~\ref{fig:parabola}
 where the FC factors depend exponentially on the vibrational numbers.
 In the limit of strong relaxation the current suppression is simply
 due to the exponentially small partial currents of the few vibrational
 states which are thermally occupied at low $T \ll \omega$.
 In the absence of relaxation the FC-factors also
 prevent the excited states from actually
 becoming occupied for $V \ll 2 \omega \lambda_1^2$.
 The resulting non-equilibrium vibrational distribution induced by the
 tunneling is ``equilibrium-like'' as was noted before~\cite{Mitra04b,Koch04b}.
 This is due a type of {\em feedback} mechanism in the tunneling
 transitions, as we will now explain.
 In principle at finite bias voltage arbitrarily high vibrational
 excitations can be accessed via {\em cascades}
 of single-electron tunneling processes.
 However, at low bias voltage
 transitions which lie outside the bias window in
 Fig.~\ref{fig:parabola} (i.e.
 $0_{q}\rightarrow 1_{q'}$ resp.
 $0_{q} \leftarrow 1_{q'}$ ) correspond to large changes of the
 vibrational energy and have exponentially larger rates.
 Once the initial states for these transitions start to be  occupied,
 the total rate for populating the lowest states
 (transitions outside the bias window) becomes much larger
 than the total rates of its decay (transitions inside bias window).
 As a result only the low-lying states are occupied due
 to the large asymmetry between the rates.
 Compared with the strong relaxation limit vibrational excitations are
 slightly more favored and therefore the current suppression is less
 severe~\cite{Koch04b} in this limit (not shown).
 The central observation is that although the occupations decrease
 strongly with vibrational number this is compensated by the
 exponential increase of the partial currents.
 The {\em  main} contribution to the current comes from the excited states.
 The ``inverted'' dependence of the FC-factors on energy (vibrational
 numbers) thus stabilizes the lowest vibrational state and enhances the
 sensitivity of the current to the small occupations of the vibrational
 excitations.  This is at the basis of the weak NDC and small current
 peaks which we will discuss now.
\begin{figure}
  \includegraphics[scale=0.7,angle=-90]{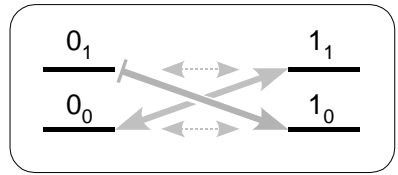}
  \caption{\label{fig:cascade_L}
     Relevant transitions for Fig.~\ref{fig:G_one_L}.
     For clarity only vibrational energy differences are indicated.
     For $\mu > \mu^1_0$ a {\em feedback} cascade of transitions keeps
     the state $1_0$ nearly fully occupied leading to NDC and current
     peaks.
}
\end{figure}
\\
{\em Weak NDC and current peak.}
 Apart from the quantitative effect on the current suppression,
 there appear interesting qualitative new features in the absence of relaxation.
 The many visible resonance lines in Fig.~\ref{fig:G_one_L} form a
 {\em Franck-Condon progression} which extends beyond voltages
 $\sim 2\omega\lambda_1^2$.
 Interestingly, for asymmetric gate energy $\mu-\mu_0^1 > \omega /2$
 the current can be reduced at the resonances
 $\mu_L=\mu_k^1,k=1,2,\ldots$
 (black features on right in Fig.~\ref{fig:G_one_L}).
 Here the transitions $0_q \rightarrow 1_{q+k},q=0,1,\ldots$ become
 allowed and one could naively expect the current to increase since
 this favors the population of the excited states
 $1_{q+k},q>0$ which are responsible for the main current
 contributions.
 Actually, the opposite happens.
 Due to the asymmetric gate-energy $\mu - \mu^1_0 > 0$ a {\em
   feedback} mechanism is active which involves a cascade of single-electron
 transitions.  This is illustrated in Fig.~\ref{fig:cascade_L}.
 In the regime of voltages where NDC occurs the charged vibrational
 ground state $1_0$ is stabilized, not only relative to the neutral
 states (due to the Coulomb blockade) but also relative to the
 vibrationally excited charged states $1_{q},q>0$.
 Due to the asymmetric gate energy
 state $1_0$ has less transitions which depopulate it
 than which populate it, see  Fig.\ref{fig:cascade_L}.
 Due to the strong increase of the FC-factors with vibrational number
 this asymmetry in the rates causes a nearly complete occupation of
 $1_0$ for asymmetric gate energies.
 Upon increasing the bias,
 at resonances where the excitations $1_{q+k}$ become accessible
 other excitations $0_{q'}$ with $q'\approx q+k$ are also
 favored via subsequent tunneling processes
 involving small changes in the vibrational number.
 These subsequent transitions are allowed already at low $V$.
 The excitations $0_{q'}$ decay with large rates back to the
 the state $1_0$ and its first few excitations, since they change the
 vibrational number by a large amount.
 Importantly, the reverse of the latter transition,
  $0_{q'} \leftarrow 1_{0}$ is not
 allowed at low $T$, see Fig.~\ref{fig:cascade_L}.
 Therefore the occupation of $1_0$ is effectively
 increased at the expense of the excited states which
 contribute most to the current and NDC occurs.
\\
 Small current peaks of width $\propto T$ occur in the intermediate
 region $0<\mu-\mu^1_0 \lesssim \omega/2$
 (white-black double lines in Fig.~\ref{fig:G_one_L} and inset of
 Fig.~\ref{fig:I_one_L}).
 These signal a redistribution of the vibrational energy
 when the bias is tuned through the resonance.
 In this case the above feedback mechanism can only become effective
 when the transition energy lies sufficiently close to $\mu_L$. Thus
 initially the current  rises but once the excited states
 become sufficiently populated they start to relax via the feedback and
 the current drops again.
\\
 Finally, Fig.~\ref{fig:I_one_L} shows that even though the
 absolute current-step amplitude exponentially increases with
 increasing bias voltage~\cite{Koch04b},
 the NDC becomes relatively less pronounced.
 Careful inspection reveals that at sufficiently large gate energy
 $|\mu-\mu^1_0|$ the resonances initially correspond to current drops
 but with increasing voltage turn into peaks and finally become current steps.
 The increasing bias eventually compensates for the gate-asymmetry and
 the feedback mechanism becomes ineffective.
\\
 In the limit of strong relaxation at low $T \ll \omega$
 only transitions inside the bias window along the $q=0$ and $q'=0$
 axis in Fig.~\ref{fig:parabola} play a role (cf. Eq.~\ref{eq:rate_eq}).
 The charged vibrational ground state is stabilized
 due to a gate voltage $\mu-\mu_0^1>0$ and the strong relaxation.
 However, no NDC or current peaks can occur at the resonances discussed above
 since the vibrational distribution is not affected by the tunneling
 in this limit. The feedback mechanism is cut off:
 after each single-electron tunneling process the excitation relaxes
 on a much shorter time scale and the next tunneling process starts
 from a vibrational ground-state again.
 Indeed one can show explicitly that in the limit of complete
 relaxation the NDC and the current peaks disappear
 in the one-orbital model for arbitrary spin and orbital degeneracy
 (see Appendix~\ref{app:relax}).
 We note that at resonance lines $\mu_R=\mu_{-k}^1$ for
 $k=1,2,\ldots$ the current always increases, independent of
 the relaxation.
 Here the transitions $1_q \rightarrow 0_{q+k}$ become
 allowed whereby $1_0$ can decay and repopulate the excited states
 which carry the current.
\subsection{\label{sec:strong_ground}
Strongly coupled ground state
}
 We now demonstrate that in a two-orbital model with asymmetric
 coupling to the vibration already at moderate coupling a new but weak
 NDC effect occurs which is robust against strong relaxation,
 in contrast to the weak NDC in the one-orbital model.
 Additionally, the non-equilibrium feedback mechanism responsible for
 the weak NDC in the one-orbital model leads to strong NDC effects for
 two competing orbitals.
 These general statements carry over to the case of multiple
 non-degenerate orbitals with asymmetric coupling occupied by at most
 one electron due to Coulomb blockade.
 We start our discussion with an intermediate case where only the weak
 NDC occurs.
\subsubsection{$\lambda_{1}^2 \gtrsim 1 \gg \lambda_{2}^2$ Weak NDC - Current oscillations
\label{sec:lS}}
\begin{figure}
  \includegraphics[scale=0.7]{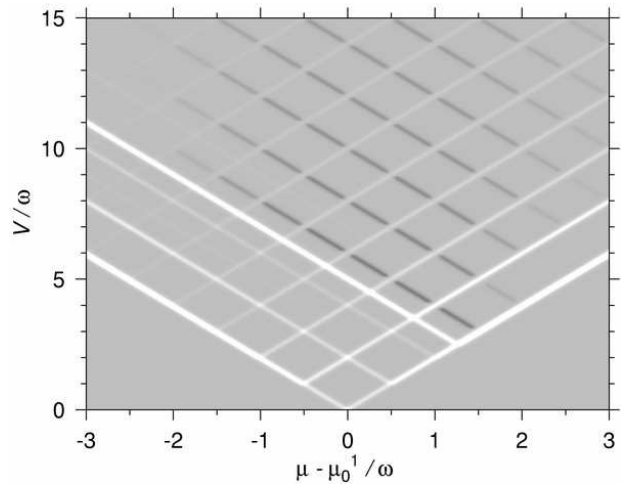}
  \caption{\label{fig:G_lSl}
    Differential conductance $dI/dV(\mu,V)$
    for $\lambda_1=1.4$, $\lambda_2=0.1$, $\Delta/\omega=2.5$.
    Current steps (white lines with negative slope) due to
    stronger coupled orbital~1
    become current {\em drops} (dark lines) as soon as the weakly
    coupled orbital~2 contributes to the current.
  }
\end{figure}

\begin{figure}
  \includegraphics[scale=0.33,angle=-90]{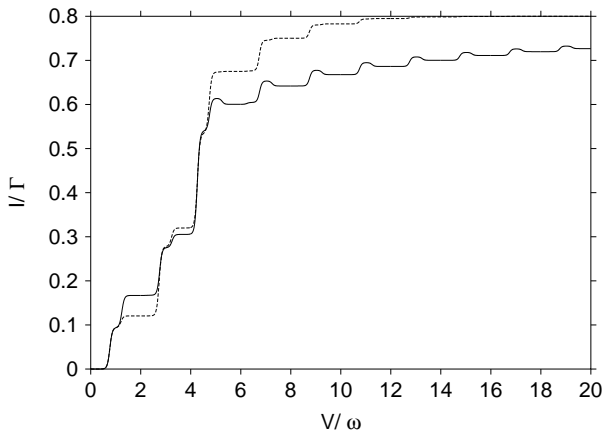}
  \caption{\label{fig:I_lSl}
    Current-voltage characteristic for $\mu-\mu^1_0= 0.375\omega$
    in Fig.~\ref{fig:G_lSl}
    for non-equilibrium (full, Eq.~(\ref{eq:I})) and equilibrium
    vibrations (dashed, Eq.~(\ref{eq:Ieq})).
    The oscillatory pattern of current steps and drops sets in
    only when orbital~2 contributes to the current (here $V>5.0 \omega$).
  }
\end{figure}

 The differential conductance and typical I-V curve in
 Fig.~\ref{fig:G_lSl} and~\ref{fig:I_lSl}, respectively,
  display a number of features which can
 be understood by considering the two orbitals invidually.
 Vibrational excitations of the stronger coupled
 state~1 form a FC progression of resonances
 far beyond $V \approx 2 \lambda_1^2 \omega$.
 In contrast, the excitations of the weakly coupled orbital~2 only
 show up as a single $dI/dV$ peak at $\mu_L=\mu^{2}_0$.
 The resonances $\mu_L=\mu^1_k$ of the stronger coupled state~1,
 correspond to current steps for $k \leqslant [\Delta/\omega]$.
 Interestingly,
 once the weakly coupled state 2 has started to contribute
 to the current, $\mu_L> \mu^2_0$,
 these turn into {\em anti-resonances}, $k > [\Delta/\omega]$,
 (dark lines in Fig.~\ref{fig:G_lSl} with negative slope).
 The resonances, $\mu_R=\mu^1_{-k},k \geq 0$, always correspond to
 current steps (white lines in Fig.~\ref{fig:G_lSl} with positive
 slope).
 A distinctive feature is that the current steps and drops due to the
 excitations of the stronger coupled state have opposite gate voltage
 dependence. Since they are of the same order of magnitude they give
 rise to current {\em oscillations} on the slowly saturating
 background.
 We note that for identical parameters the one-orbital model
 (i.e. with either $\lambda_1=0.1$ or $1.4$) produces negligible NDC
 effects.
 The origin of the enhancement of NDC is the following.
 For moderate gate energy $\mu-\mu^1_0$ the partial currents of states
 $1_q$ are small compared with those of states $2_q$ and $0_q$,
 cf. Fig.~\ref{fig:parabola}.
 At resonance lines $\mu_L=\mu^1_{k}$ the transitions
 $ 1_{q+k} \leftarrow 0_{q}$ become allowed, and the occupations of
 the states with the larger partial currents are reduced
 due to the Coulomb correlation between the two orbitals.
 The asymmetry between the partial currents is only present when the
 weakly coupled orbital~2 is accessible,
therefore the current drops at these resonances only once both
 orbitals are accessible, i.e. $\mu_L  \geqslant \mu^2_0 \geqslant \mu_R$.
 In contrast, the resonances $\mu_R=\mu^1_{-k},k>0$ correspond to current steps
 since here states $1_q$ are depopulated in favor of the states with
 larger partial currents.
\\
 Although the current oscillations in Fig.~\ref{fig:G_lSl} seem very
 similar to those for the one-orbital model in Fig.~\ref{fig:G_one_L},
 there is an important difference:
 here the NDC is not completely suppressed in the strong relaxation limit,
 although larger values of the dominant coupling $\lambda_1$ are
 required for visibility comparable with the limit of no relaxation.
 In Appendix~\ref{app:relax} we prove that in the strong relaxation
 limit a drop of the current can only occur along resonance lines
 $\mu_L = \mu^i_{k}$ (negative slope) if it occurs.
 A condition for the visibility of NDC is that the total current below the
 resonance is larger than compared with the partial current of the
 orbital causing it.
 This requires a weakly coupled orbital, $\lambda_2^2 \ll 1$, with
 large partial currents in combination with a stronger coupled orbital,
 $\lambda_1^2 \gtrsim 1$.
 These are roughly the same conditions as for the
 visibility of the oscillating current in the non-equilibrium
 limit.
 However, we point out that under identical bias and gate voltage
 conditions the NDC need not to be visible in both the non-equilibrium
 and equilibrium limit, see for instance Fig.~\ref{fig:I_lSl} and
 Fig.~\ref{fig:I_LSs} below.
\\
 In summary, the current oscillation occurs due to the competition
 between transport channels with significantly different partial
 currents.
 It is a {\em Coulomb repulsion} effect: the current calculated
 without the effective correlations ($v=0$), always increases with
 $V$ (see Appendix~\ref{app:uncorr}).
\subsubsection{ $\lambda_{1}^2 \gg 1 \gg \lambda_{2}^2$ Strong NDC
  \label{sec:LS}}
 \begin{figure}
   \includegraphics[scale=0.7]{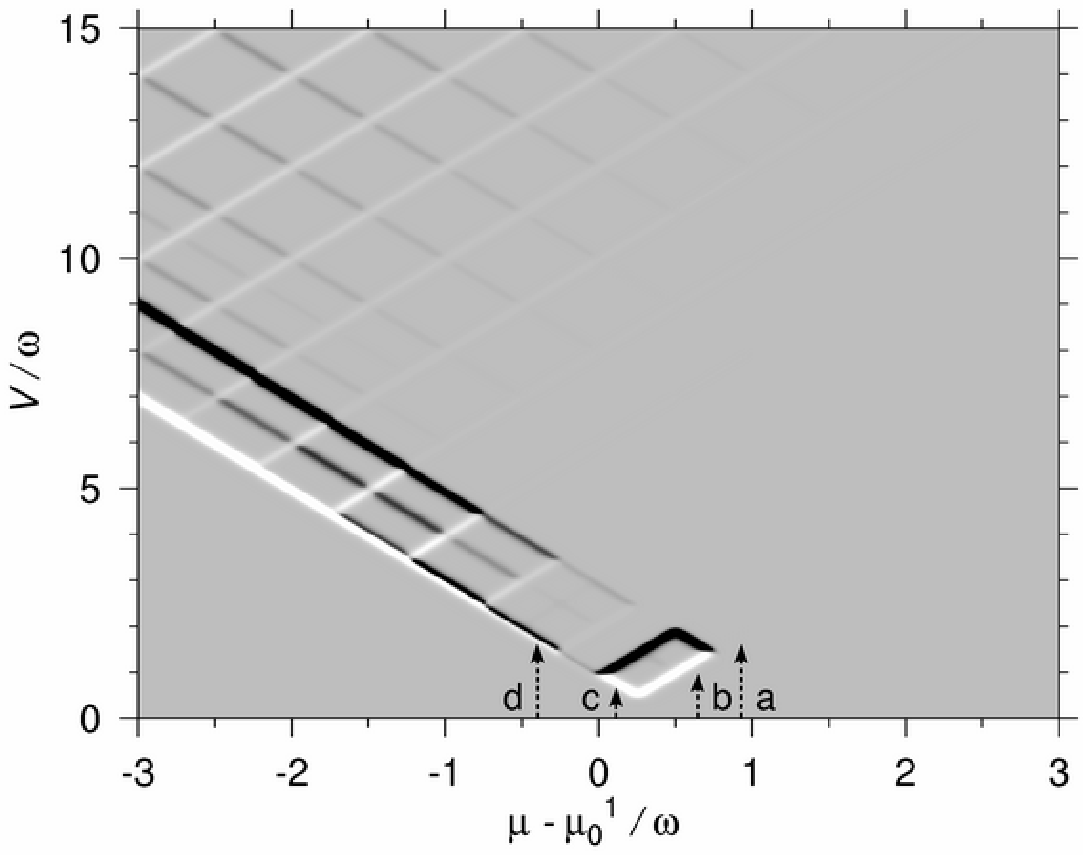}
   \caption{\label{fig:G_LSs}
     $dI/dV(\mu,V)$
     for $\lambda_1=5$, $\lambda_2=0.1$ and $\Delta/\omega=0.5$.
     The isolate region at low bias is due to the small excitation
     energy $\Delta < \omega$ and disappears for $\Delta > \omega$.
     The current is strongly suppressed once the {\em weakly
       coupled} state~2 can be excited,
     $\mu_L > \mu^2_1$ and $\mu_R < \mu^2_{-1}$ (thick black lines).
     A current {\em peak} (white-black double line) appears when
     this orbital becomes accessible at $\mu_L=\mu^2_0$.
   }
   \includegraphics[scale=0.4,angle=-90]{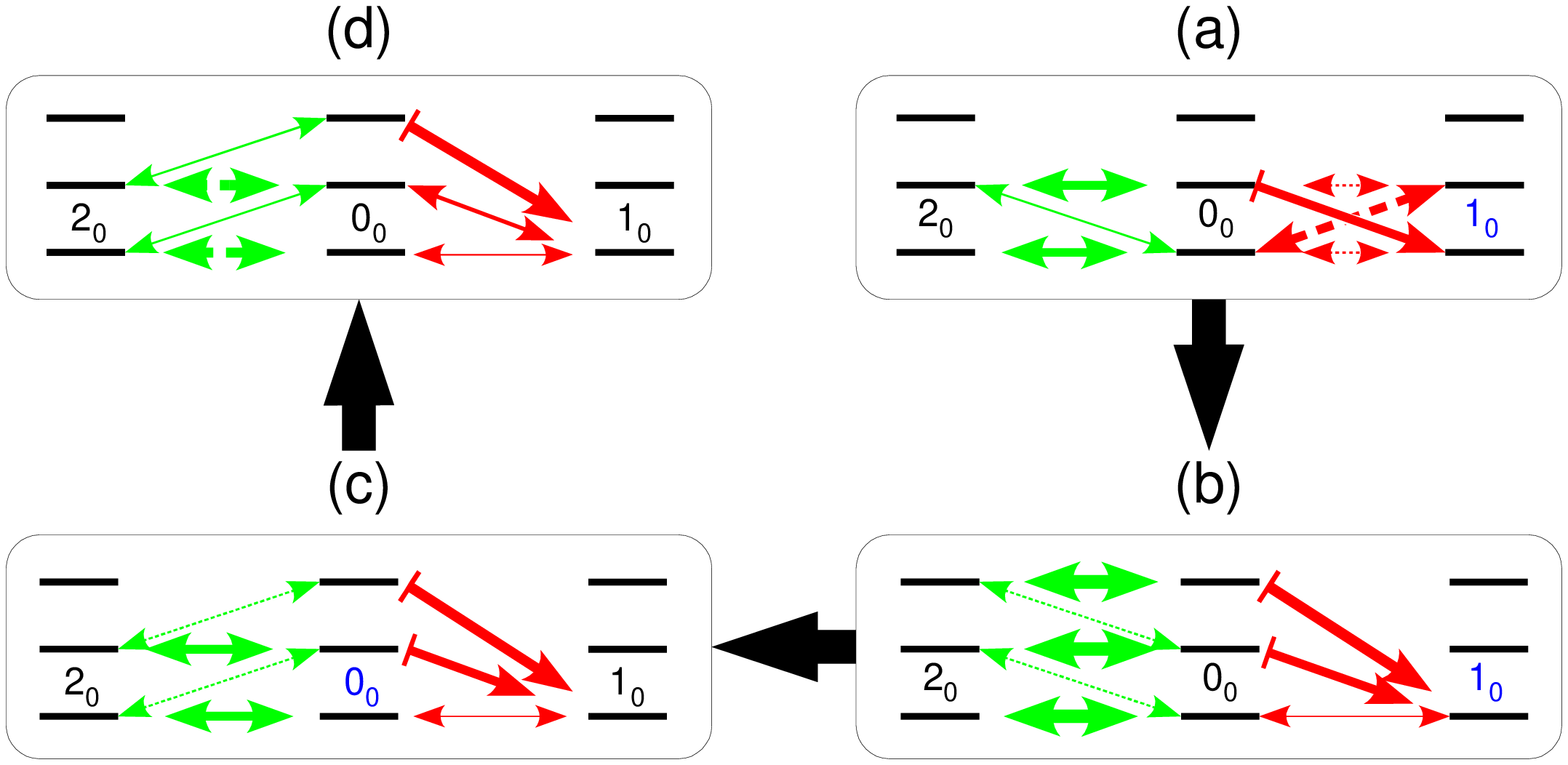}
   \caption{\label{fig:cascade_LSs}
     Relevant transitions in low bias section of Fig.~\ref{fig:G_LSs}.
     Indicate are transitions which cause NDC to occur when increasing
     the bias along (a)-(d).
     For clarity only vibrational energies are indicated.
     Arrow thicknesses indicate relative magnitude of the transition rates
     between vibrational excitations of a pair of states $0_q
     \leftrightarrows i_{q'},q,q'=0,1,2,\ldots$ where $i=1$ or $i=2$.
     Rates between different pairs of states are of different order of
     magnitude since different couplings ($\lambda_1$ or $\lambda_2$) is
     involved.
     The cascades of transitions give rise to a {\em feedback} into the lowest
     vibrational state $1_0$ of strongly of the coupled orbital.
   }
 \end{figure}
\begin{figure}
  \includegraphics[scale=0.33,angle=-90]{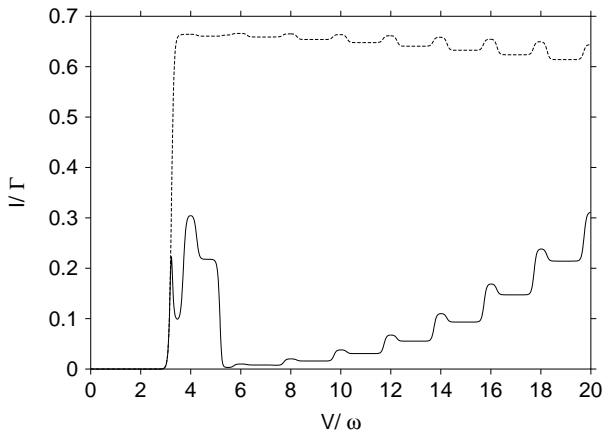}
  \caption{\label{fig:I_LSs_peak}
    $I(V)$ for $\mu-\mu^1_0=-1.125\omega$ in Fig.~\ref{fig:G_LSs} [case (d)]
    for non-equilibrium (full, Eq.~(\ref{eq:I})) and equilibrium vibrations
    (dashed, Eq.~(\ref{eq:Ieq})).
    The current sets on with a {\em peak} of width $\propto T$, the other
    local maxima are thermally smeared plateaus.
    The downward trend of the equilibrium curve only applies at low
    voltages $V < 2 \lambda_1^2\omega$, it saturates to the ``electronic
    limit'' at larger voltages.
  }
 \end{figure}

 When the charged ground state couples strongly to the vibration ,
 $\lambda_1^2 \gg 1$, and the electronic excitation lies low,
 $\Delta < \omega$,
 we find the typical structure of Fig.~\ref{fig:G_LSs}.
 The current oscillations are clearly visible again and this
 set of resonances needs no further discussion.
 An obvious difference with Fig.~\ref{fig:G_lSl} is the finite gap
 $\geq \Delta$ for any gate voltage.
 In fact, in addition to the Coulomb blockade regime where
 $\mu_L < \mu^1_0$ or $\mu^1_0 < \mu_R$ resp., 
 the current is suppressed in the entire
 strip $\mu^2_0 \geq \mu_L \geq \mu^1_0 , \mu_R > \mu^1_0$ (i.e. where
 the excited orbital~2 is not yet accessible).
 This is due to the exponentially small FC-factors $F^1_{q q'},q,q'\ll
 \lambda_1^2$ of the lowest orbital which now couples strongly to the
 vibration (cf. Sec.~\ref{sec:single}).
 When increasing the bias voltage above the gap the resulting current
 depends strongly on the order in which additional transitions become
 allowed i.e. on the gate voltage.
 Basically four different situations can occur which are labeled
 (a)-(d) in Fig.~\ref{fig:G_LSs} and the relevant transitions are
 schematically indicated in Fig.~\ref{fig:cascade_LSs}.
\\
 {\em (a) Stabilization of the charged ground state.}
 In the upper-right region in Fig.~\ref{fig:G_LSs},
 the current may be expected to flow:
 the transitions  $2_{q'} \leftrightarrows 0_q$ are
 energetically allowed for at least $q=q'=0$, for which the FC-factor
 is large, $F^2_{00} \approx 1$.
 Instead the current is strongly suppressed and increases in small
 steps of increasing height with increasing bias.
 This is  very similar to the situation in Fig.~\ref{fig:I_one_L}
 where the transport is dominated by
 a single  orbital with $\lambda_1^2 \gg 1$.
 Between states $1_{q'}$ and $0_q$ the feedback mechanism discussed in
 Sec.~\ref{sec:single} is operative
 which keeps state~$1_0$ almost fully occupied with increasing bias,
 see Fig.~\ref{fig:cascade_LSs}(a) and compare with Fig.~\ref{fig:cascade_L}.
 The presence of orbital~2 further enhances the feedback since the
 states $0_q$ (which feed back into $1_0$) can now also be populated
 via cascades of transitions involving orbital~2.
\\
 \begin{figure}
   \includegraphics[scale=0.33,angle=-90]{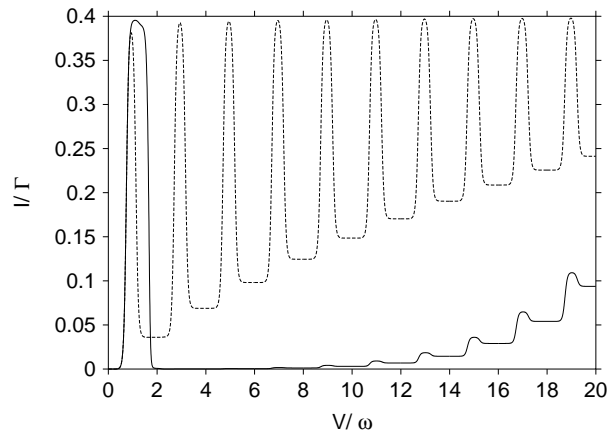}
   \caption{\label{fig:I_LSs}
     $I(V)$ for $\mu-\mu^1_0=0.375\omega$ in Fig.~\ref{fig:G_LSs} [case
     (b)],
     for non-equilibrium (full, Eq.~(\ref{eq:I})) and equilibrium vibrations
     (dashed, Eq.~(\ref{eq:Ieq})).
     The isolated current plateau at low bias $V \leq 2 \omega$ and the
     subsequent current suppression are non-equilibrium effects.
     For this gate energy both the current oscillations
     and the rising background current are {\em enhanced} when one allows for
     strong relaxation (in contrast to the one-orbital model).
   }
 \end{figure}
 {\em (b,c) Isolated region.}
 The small diamond-shaped region in Fig.~\ref{fig:G_LSs} at {\em
 finite voltage} $\Delta < V < 2 \omega$ is remarkable.
 Inside it the current is non-zero and beyond any of its four defining
 boundaries in the plane of gate- and bias voltage
 the current is completely suppressed.
 A typical $I(V)$ curve through this region is shown in Fig.~\ref{fig:I_LSs}.
 The fact that the region is bounded by a {\em NDC line with positive slope}
 proves that it must be caused by non-equilibrium vibrational
 effects (Sec.~\ref{sec:model}),
 see also the strong relaxation result in Fig.~\ref{fig:I_LSs}.
 This region can be reproduced by truncating the spectrum to 5
 states: $0_q,2_q, q=0,1$ and $1_0$, see also
 Fig.~\ref{fig:cascade_LSs}.
 At the low bias side, Coulomb blockade and the small FC factor $F^1_{00}$
 discussed above are responsible for the current suppression.
 Inside this region only the transitions
 $2_0 \leftrightarrows 0_0 \leftrightarrows 1_0$ are allowed.
 The stationary occupations follow from Eq.~(\ref{eq:dotP}):
 $P^0_0=1/5, P^1_0= P^2_0=2/5$ and the current is:
 \begin{eqnarray}
   I \approx  \frac{1}{2}I_0^2 P^2_0+I_0^0 P^0_0=\frac{2}{5} \Gamma
 \end{eqnarray}
 This is less than $2\Gamma/3$ and $4\Gamma/5$, the maximal current through
 one and two orbitals (without the vibration), respectively,
 due to the partial occupation of the strongly coupled state~$1_0$
 with suppressed partial currents.
 Note that state $1_0$ is not yet blocking the transport.
 At the high bias side, the current becomes suppressed
 when the first neutral excited state $0_1$ can be reached either via
 the cascade
 $0_0 \rightarrow 2_1 \rightarrow 0_1$
 (NDC line with negative slope, case (b) in Figs.~\ref{fig:G_LSs} and \ref{fig:cascade_LSs})
 or
 $2_0 \rightarrow 0_1$
 (NDC line with positive slope, case (c)).
 State~$0_1$ can decay to $1_0$ when an electron enters the molecule
 through {\em either} junction i.e. the reverse transition
 $0_1 \leftarrow 1_0$ is suppressed at low temperature
 $T \ll \omega-\Delta$.
 Now state $1_0$ is almost fully occupied since it is populated
 much faster than it can decay: it is blocking the transport since
 the current it limited by the very small sum of its decay rates.
 The feedback loop thus involves both the weakly and
 strongly coupled state:
 the competition between the two orbitals which couple asymmetrically
 to the vibration causes the NDC to be much stronger
 compared with the one-orbital case (Sec.\ref{sec:single}).
 From the above it follows that the diamond-shaped region disappears
 for excitation energy $\Delta > \omega$. However, the feature (a)
 and (d), which we discuss now, will still be present.
\\
 {\em (d) Current peak.}
 At the resonance line $\mu_L=\mu^{2}_{0}$ where the weakly coupled
 orbital~2 starts to participate a single large current step
 could be expected.
 Remarkably, NDC occurs {\em in the middle} of this resonance,
 producing a {\em current peak} (white-black double line
 with negative slope in Fig.~\ref{fig:G_LSs})
 whose width is proportional to the electron temperature $T$.
 This sharp features stands out between the thermally broadened
 plateaus in Fig.~\ref{fig:I_LSs_peak}.
 The origin of the peak is a strong competition between the charged
 states $1_0$ and $2_0$ in the narrow energy window
 $| \mu_L-\mu^2_0 | \sim T$, involving the feedback via neutral excited states $0_q$.
 This is most simply illustrated by considering gate energies
 $ \Delta/2-\omega < \mu-\mu^1_0 < \Delta/2-\omega/2$
 where a minimal set of 6 states is sufficient to understand the peak,
 see Fig.~\ref{fig:cascade_LSs}(d).
 At the rising side of the peak the rate of the transition
 $2_0 \leftarrow 0_0$
 (through the $V$ dependence in the Fermi-function,
 cf. Eq.~\ref{eq:rate})
 has increased sufficiently to enhance the current relative to the
 very small value supported only by the transitions to / from the
 blocking state, $1_0 \leftrightarrows 0_0$.
 Due to the gate energy the simplest feedback loop involves a cascade
 of 6 transitions:
 $0_0 \rightarrow 2_0 \rightarrow
  0_1 \rightarrow 2_1 \rightarrow
  0_2 \rightarrow 1_0$.
 This feedback initially remains ineffective since the
 excited state $0_2$ is not yet sufficiently populated.
 The vibrational distribution is equilibrium-like and the current
 follows the result for equilibrated vibrations, see
 Fig.~\ref{fig:I_LSs_peak}.
 As one increases $V$ the occupations of the excited states in the
 feedback loop increases
 and a maximum is reached. Here 
 the feedback dynamically starts to trap the molecule in
 the state $1_0$.
 The occupations of the excited states and the current now start to
 decrease and reach a lower value (although higher than before the peak).
 The current peak thus signals this redistribution of the
 vibrational energy in a small bias window.
 We note that the current is not completely suppressed: this only
 happens beyond the second strong NDC line $\mu_L=\mu^{2}_{1}$
 (cf. Fig.~\ref{fig:I_LSs_peak}) as may be understood by considering
 Fig.~\ref{fig:cascade_LSs}(d).
 The peak can thus be considered a precursor of the full onset of the
 feedback mechanism.
 For lower $ \mu-\mu^1_0 < \Delta/2-\omega$ a similar argument
 involving more than 6 states explains why the peak becomes a step and
 why simultaneously the strong NDC along $\mu_L = \mu^2_1$ is further
 enhanced.
\\
 {\em Intermediate relaxation.}
 Upon increasing the vibrational relaxation rate $\gamma$
 (cf. Sec.~\ref{sec:model}) starting from zero,
  the strong relaxation result is approached as expected.
 The relaxation cuts off the cascade of transitions leading to the
 blocking state and reduces the importance of the feedback for the
 transport.
 Now the NDC becomes more pronounced at resonances $\mu_L=\mu_k^1,
 k=1,2,...$ where the transitions $0_{q}\rightarrow 1_{q+k}$ become
 allowed.
 These enhance the occupation of $1_0$ due to relaxation and suppress
 the current.
 However, this approach is rather slow at low bias voltages:
 the NDC lines marking the isolated region remain clearly visible.
\\
 In summary, the strong NDC lines $\mu_L=\mu^2_1$ and
 $\mu_R=\mu^2_{-1}$ are associated with excitations of the state with
 {\em weakly} coupled to the vibration (state~2).
 However, the {\em strongly} coupled state~1 is actually blocking the transport.
 The weakly coupled state allows an excess vibrational energy to
 accumulate on the molecule (through a cascade of tunneling processes)
 which is subsequently spent to trap the
 molecule in the strongly coupled state (in a single tunneling process).
 The blocking state can thus be reached under very general energetic
 conditions. Therefore NDC effects become strong when 2 (or more)
 orbitals which couple asymmetrically to the vibration compete in the
 transport.
 Finally we note that Fig.~\ref{fig:G_LSs} is reminiscent of the
 signatures of spin-blockade of tunneling. There the resonance line
 marking the transition between ground states can be terminated at
 finite bias and the current is only recovered when excited states
 become accessible~\cite{Weinmann95}.
 The resonance line thus shows a kink.
 Here the kink in the resonance line is more drastic since also
 transitions to {\em many} vibrationally excited states are suppressed.
 Such details are of importance to distinguish NDC due to spin
 excitations in molecules~\cite{Romeike04} from
 effects due to vibrational excitations.
 \subsection{\label{sec:strong_excited}
   Stronger coupled excited state
 }
 We now consider the case opposite to Section~\ref{sec:strong_ground}
 where the excited orbital~2 takes up the role of the {blocking}
 orbital due to a strong(er) coupling to the vibration and the lower
 orbital~1 is weakly coupled.
 Vibration-assisted tunneling processes now stabilize the electronically
 excited state of the charged molecule i.e. an {\em excess charge and
   energy} may be stored on the molecule by the feedback mechanism.
 We focus on the resulting qualitative differences with respect to
 Sec.~\ref{sec:strong_ground} resulting from this population-inversion
 controlled by the bias voltage.
 As a simple reference point, we start again from an intermediate case.
 \subsubsection{$\lambda_{1}^2 \ll 1 \lesssim \lambda_{2}^2$
   Weak NDC - current oscillations
   \label{sec:Sl}}
\begin{figure}
   \includegraphics[scale=0.7]{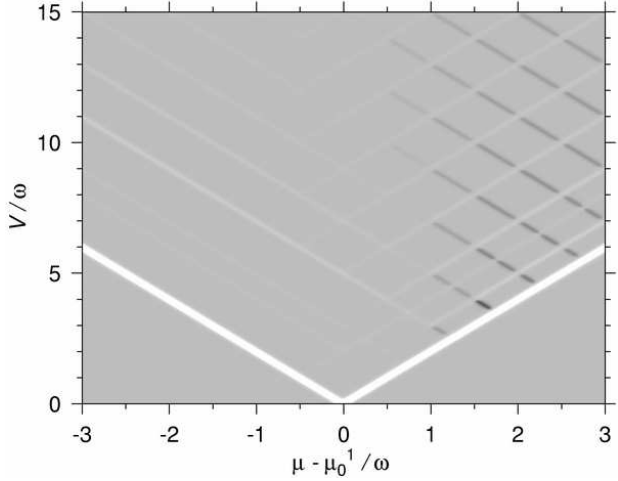}
   \caption{\label{fig:G_Sll}
     $dI/dV(\mu,V)$ for
     $\lambda_1=0.1$, $\lambda_2=1.4$, and $\Delta/\omega=2.5$.
     The oscillations are similar to those in Fig.\ref{fig:G_Sll}.
     The current drops (black lines with negative slope) are again
     due to the {\em stronger} coupled orbital~2.
   }
 \end{figure}

 Fig.~\ref{fig:G_Sll} shows the typical differential conductance
 for $\Delta > \omega$.
 Similar to Fig.~\ref{fig:G_lSl} the weakly coupled state (state~1)
 basically shows up as one large current step and the many resonance lines
 correspond to the excitations of the strongly coupled state (here
 state~2).
 However, at the resonances $\mu_L =\mu^2_{k},k
 \geq [\Delta/\omega]$ we have current drops for small $|\mu -
 \mu^2_0|$ (i.e. away from the charge degeneracy point)
 and current steps for large $|\mu - \mu^2_0|$.
 Therefore, around the charge degeneracy point 
 $\mu=\mu^1_0$ the first few excitations beyond the
 electronic excited line $\mu_L = \mu^2_0$ show up as current steps
 and only at higher bias voltage turn into current drops,
 in contrast to the case where the lower orbital couples stronger to
 the vibration (Fig.~\ref{fig:G_lSl}).
 Up to now we have only found features in the differential conductance
 (either positive or negative) when incoming electrons excite the
 vibration with their excess energy. This is expected at low
 temperatures $T \ll \omega$.
 Interestingly, in Fig.~\ref{fig:G_Sll} a small current step at the
 resonance $\mu_L =\mu^2_{-1}$ with negative slope is visible which
 corresponds to {\em absorption} of the vibrational energy by an {\em
 incoming} electron despite the low temperature $T \ll \omega$ and
 moderate bias.
 The slightly enhanced current may be understood as an effect of
 significant heating of the molecule by the vibration assisted-tunneling.
\subsubsection{$\lambda_{1}^2 \ll 1 \ll \lambda_{2}^2$
  Strong NDC
\label{sec:SL}}
 \begin{figure}
   \includegraphics[scale=0.7]{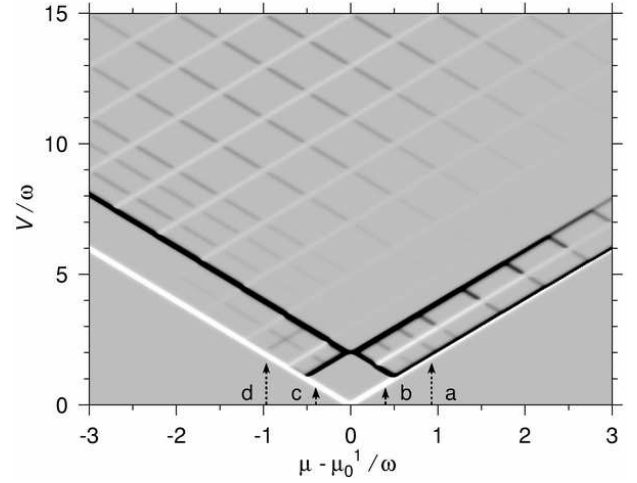}
   \caption{\label{fig:G_SLl}
     $dI/dV(\mu,V)$ for
     $\lambda_1=0.1$, $\lambda_2=5.0$, and $\Delta/\omega=2.5$.
     Similar to Fig.~\ref{fig:G_LSs}
     a two strong NDC lines define a low bias structure and
     a current {\em peak}, which now occurs at $\mu > \mu^1_0$.
     The progression of NDC lines extending up to large bias voltage
     is again due to the strongly coupled {\em excited} orbital~2.
     Remarkably, this NDC already sets in at low $V$, before this orbital
     can be directly accessed: this signals {\em absorption} of vibrational
     energy by {\em incoming} electrons.
   }
 \end{figure}
 When the coupling to the charged excited state
 becomes strong, $\lambda_2^2 \gg 1$
 the case of most interest is that of a higher excited orbital,
 $\omega < \Delta \ll \lambda_2^2 \omega$.
 The typical structure is shown in Fig.~\ref{fig:G_SLl}.
 \begin{figure}
   \includegraphics[scale=0.4,angle=-90]{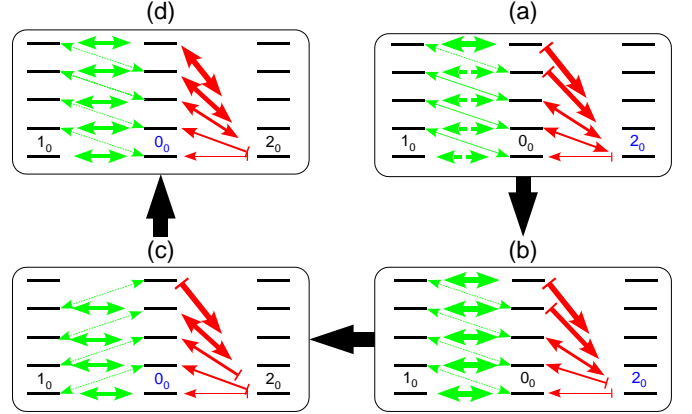}
   \caption{\label{fig:cascade_SLl}
     Same diagram as Fig.~\ref{fig:cascade_LSs} but now for low-bias
     section of Fig.~\ref{fig:G_SLl}.
     The feedback mechanism now involves more excitations $0_q,1_q,q
     \leq \Delta/\omega > 1$.
     Note the state which is stabilized by this non-equilibrium mechanism
     , $2_0$, is an {\em electronically} excited state i.e. we have a
     population inversion.
   }
 \end{figure}
 In contrast to Fig.~\ref{fig:G_LSs} we do not have a gap here
 since the change in the nuclear configuration of the two ground states is now
 small:
 the current starts to flow at the edges of the Coulomb blockade
 region $\mu_L = \mu^{1}_0$ and $ \mu^{1}_0 = \mu_R$.
 Two NDC lines stand out in Fig.~\ref{fig:G_SLl} (dark) where the current
 is significantly suppressed:
 $\mu_L = \mu^1_1$ and $\mu_R = \mu^1_{-1}$.
 Again, the appearance of the latter NDC line with positive slope is a
 proof  that non-equilibrium vibrations play a role (Sect.~\ref{sec:model}).
 At these lines the transitions  $1_{q+1} \leftarrow 0_q$ and
 $1_q \rightarrow 0_{q+1}$ between the neutral and the {\em
 weakly} coupled charged state become allowed
 which enhance the occupation of the excited states $0_q$.
 The latter feed back to the strongly coupled state $2_0$
 with rates which increase exponentially with $q$
 and the electronic excited orbital is predominantly occupied.
 We thus have a bias-controlled {\em population inversion}
 between ground- and excited state of the charged molecule
 due to their asymmetric coupling to the vibration.
 We discuss the four different situations labeled (a)-(d)
 in Fig.~\ref{fig:G_SLl} for which the relevant transitions are
 schematically indicated in Fig.~\ref{fig:cascade_SLl}.
\\
{\em (a) Current peak.}
 A current {\em peak} appears along the line
 $\mu_R = \mu^1_0$ for $\mu -\mu^1_0>\omega/2$
 (white-black double line on the right in Fig.~\ref{fig:G_SLl}).
 Compared with Fig.~\ref{fig:G_LSs} this peak occurs on the opposite
 side and remains visible up to much higher values of $\mu - \mu_0^1$.
 The mechanism causing the current peak is analogous to that discussed
 in Section \ref{sec:LS} with the roles of 1 and 2 interchanged.
 However, more vibrational excitations are involved in the
 feedback mechanism depicted in Fig.~\ref{fig:cascade_SLl} since
 $\Delta > \omega$.
 For moderate gate energy $\mu-\mu^1_0$ the current peak is a
 precursor to the population inversion:
 the feedback is only completely activated beyond the resonance above it,
 $\mu_R=\mu^1_{-1}$. There the transitions $1_q \rightarrow 0_{q+1}$
 become allowed and
 the strongly coupled state $2_0$ is predominantly occupied
 suppressing the current  (upper-right region in
 Fig.~\ref{fig:G_SLl}).
 For large asymmetric gate energy $\mu-\mu^1_0 \gg \omega /2$
 the peak actually {\em marks} the population inversion:
 the NDC at the peak gains in amplitude at the expense of the NDC
 above it at $\mu^1_{1} = \mu_R$ (opposite to Fig.~\ref{fig:G_LSs}).
 In this case state $2_0$ can already be reached by many feedback cascades
 $0_{0} \rightarrow 1_{k} \rightarrow 2_0,
  1 \leq k \leq [(\mu-\mu^1_0)/\omega]$,
 once the escape from orbital~1 ($0_0 \leftarrow 1_0$) becomes
 possible. Therefore the population inversion and current suppression
 are complete at the peak.
 For a low-lying excited orbital, $\Delta < \omega$, this is in fact
 the general situation since the cascade of transitions involved in
 the feedback is shorter.
\\
{ \em (b,c) Isolated region.}
 For $|\mu-\mu^1_0 | < \omega/2$ and $V< 2 \omega$ we have an isolated
 region in the sense that the current reaches a local maximum value
 (diamond shaped region at bottom of Fig.~\ref{fig:G_SLl}).
 This region does not occur for $\Delta < \omega$ (opposite to
 Sec.~\ref{sec:LS} where $\Delta > \omega$ suppresses the isolated region).
 Within this region only the transitions $1_0\leftrightarrows 0_0$ are allowed, 
 $P^0_0=1/3,
  P^1_0=2/3$,
 and the current
 \begin{eqnarray}
   \label{eq:Iregion_SLl}
   I \approx I^0_0 P^0_0 + \frac{1}{2} I^1_0 P^1_0 = \frac{2}{3} \Gamma
 \end{eqnarray}
 equals the maximum current which a single orbital (without the vibration) can
 carry.
 When going along (b) and (c) in Fig.~\ref{fig:G_SLl}
 we next cross the resonance lines $\mu_L=\mu_1^1$
 and $\mu_R=\mu^1_{-1}$ respectively and the current decreases as
 discussed above.
 The current suppression is complete
 once both transitions $0_q \rightarrow 1_{q+1}$ and $1_q \rightarrow
 0_{q+1}$ are allowed.
 \\
 \begin{figure}
   \includegraphics[scale=0.33,angle=-90]{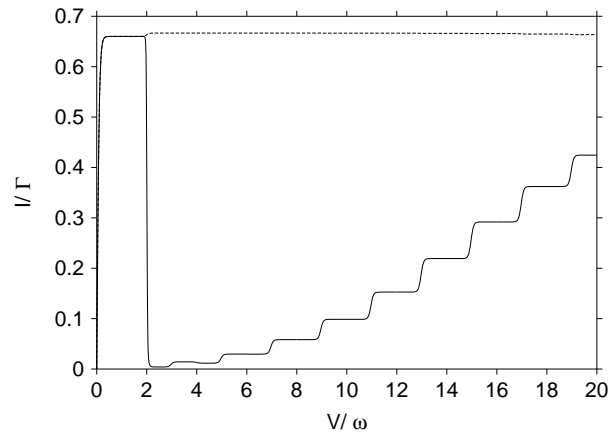}
   \caption{\label{fig:I_SLl}
     $I(V)$ for $\mu-\mu^1_0=0.0$ in Fig.~\ref{fig:G_SLl}
     for non-equilibrium (full, Eq.~(\ref{eq:I})) and equilibrium vibrations
     (dashed, Eq.~(\ref{eq:Ieq})).
     The current plateau is caused by the weakly coupled ground state~1.
     The current suppression is due to the strongly coupled excited state~2 at
     $\Delta=2.5\omega$ which is only
     accessible by a {\em single}-electron tunneling process for $V >5.0 \omega$
     (cf. Fig.~\ref{fig:G_LSs}).
     However, a cascade of such processes, whereby vibrational
     energy accumulates on the molecule, allows this state to become
     occupied already at low $V$, Figs.~\ref{fig:cascade_SLl}(b),(c).
   }
 \end{figure}
 {\em (d) Absorption by incoming electrons.}
 In Fig.~\ref{fig:G_SLl} resonances
 $\mu_L=\mu^2_{-k},k=1,2$ (negative slope) are visible which
 correspond to  absorption of the vibrational energy by an
 incoming electron ($2_{q} \leftarrow 0_{q+k}$).
 What is remarkable compared to Fig.~\ref{fig:G_Sll} is
 that the current {\em decreases} here.
 This is another signature of the population inversion
 due to the feedback mechanism.
 Vibrational energy can accumulate on the molecule through previous
 sequences of tunneling events involving only the weakly coupled
 orbital~1.
 The molecule is then ``brought to a standstill'' when, in a single
 tunneling process, an electron with an energy deficit matching the
 total accumulated vibrational energy enters: $2_0 \leftarrow 0_k$.
 For $\mu^2_k < \mu_L < \mu^2_0$ there are $[\Delta/\omega]$ such
 resonance lines where such a new trapping process becomes possible.
 Note that this can not be understood as an effect of heating since at these
 resonances the occupation of vibrational excited states is
 suppressed due to the feedback and the current is reduced.
\\
 {\em Intermediate relaxation.}
 Compared with Section~\ref{sec:LS} the non-equilibrium effects
 are more sensitive to relaxation since here roughly
 $2\Delta/\omega$ tunneling events comprise the feedback mechanism
 instead of 2 (cf.~Figs.~\ref{fig:cascade_LSs} and \ref{fig:cascade_SLl}).
 This restricts the minimal vibrational quality factor
 $\mathcal{Q}$ (writing the vibrational relaxation rate as
 $\gamma=\omega/\mathcal{Q}$,  cf. Sec.~\ref{sec:model})
 for the observation of effects due the feedback mechanism to
  $\mathcal{Q} > 2\Delta/\omega$ for $\lambda_1^2 \ll 1 \ll \lambda_2^2$.
 In contrast, for $\lambda_1^2 \gg 1 \gg \lambda_2^2$ the requirement is
 $\mathcal{Q} > 2$.
 For the cases discussed here only $\mathcal{Q} \gg 5$ is required.
 This is confirmed by our calculations for intermediate values of the
 relaxation rate $\gamma$.
 Interestingly, when increasing the vibrational relaxation rate $\gamma$
 starting from zero, the dependence of the
 amplitude NDC line with positive slope $\mu_R=\mu^1_1$ 
 (proof of a non-equilibrium distribution) is non-monotonic:
 it is initially weakened and then regains amplitude and remains clearly
 visible with increasing $\gamma$.
 Also the current peak shifts to larger values of $\mu-\mu^1_0$ and $V$ but
 remains visible.
 The low-bias effects of the excited state are however suppressed
 since cascades responsible for the population inversion effect are
 cut off by the relaxation.
 \\
 In summary, it is remarkable that in all discussed cases (a)-(d) the
 excited state $2_0$ dominates the current at low bias where it
 can not be reached from the neutral ground state $0_0$ by a
 single-electron tunneling process.
 The strong deviation from equilibrium is induced by cascades of single
 electron tunneling processes.
\subsection{
\label{sec:both}
 Non-degenerate strongly coupled states
}
 Having analyzed the above cases in detail, we can now
 restrict ourselves to a brief classification of the results
 for non-degenerate states $\Delta > 0$ where both
 $\lambda_1^2 \ne \lambda_2^2 > 1$.
 Here the feedback mechanism discussed above produces more complex
 results.
 The basic change is that when the {\em weakest} coupling is increased
 the feedback mechanism becomes less efficient at populating vibrational
 excitations of the neutral molecule (compare with the feedback mechanism
 for a single orbital Sec.~\ref{sec:single}).
 More vibrational excitations of the weakly coupled orbital and the
 neutral state must be accessible by a single tunneling process in
 order to fully activate the feedback and trap the molecule in {\em
   strongest} coupled orbital.
 The patterns of NDC lines will therefore extend over a broader range of
 applied voltages.
 \begin{figure}
   \includegraphics[scale=0.7]{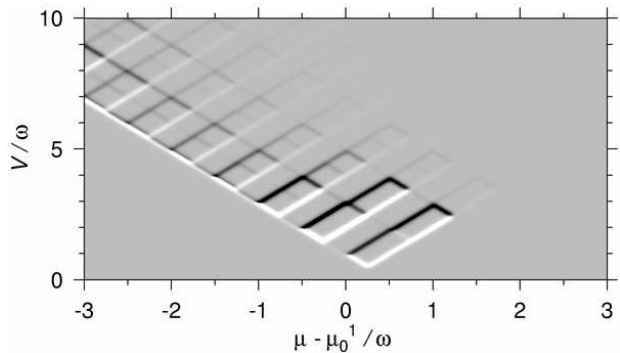}
   \caption{\label{fig:G_Lls}
     $dI/dV(\mu,V)$ for
     $\lambda_1=5.0$, $\lambda_2=1.1$ and $\Delta/\omega=0.5$.
   }
 \end{figure}
 \begin{figure}
   \includegraphics[scale=0.7]{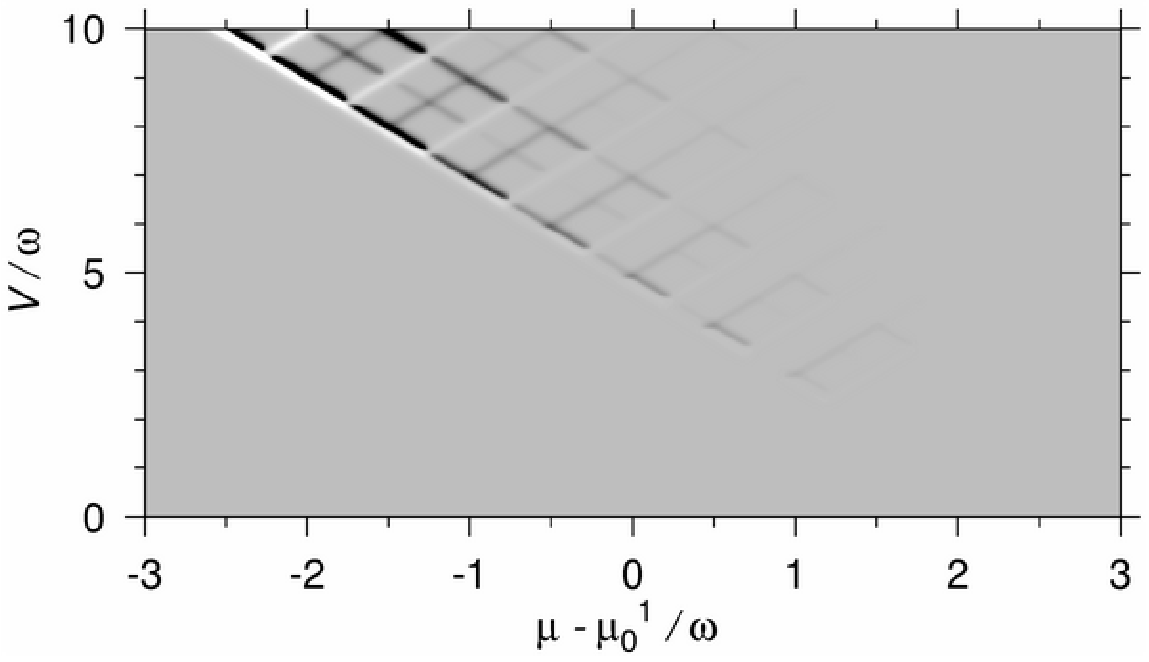}
   \caption{\label{fig:G_Lll}
     $dI/dV(\mu,V)$ for
     $\lambda_1=5.0$, $\lambda_2=1.1$ and $\Delta/\omega=2.5$.
   }
 \end{figure}
 \begin{figure}
   \includegraphics[scale=0.7]{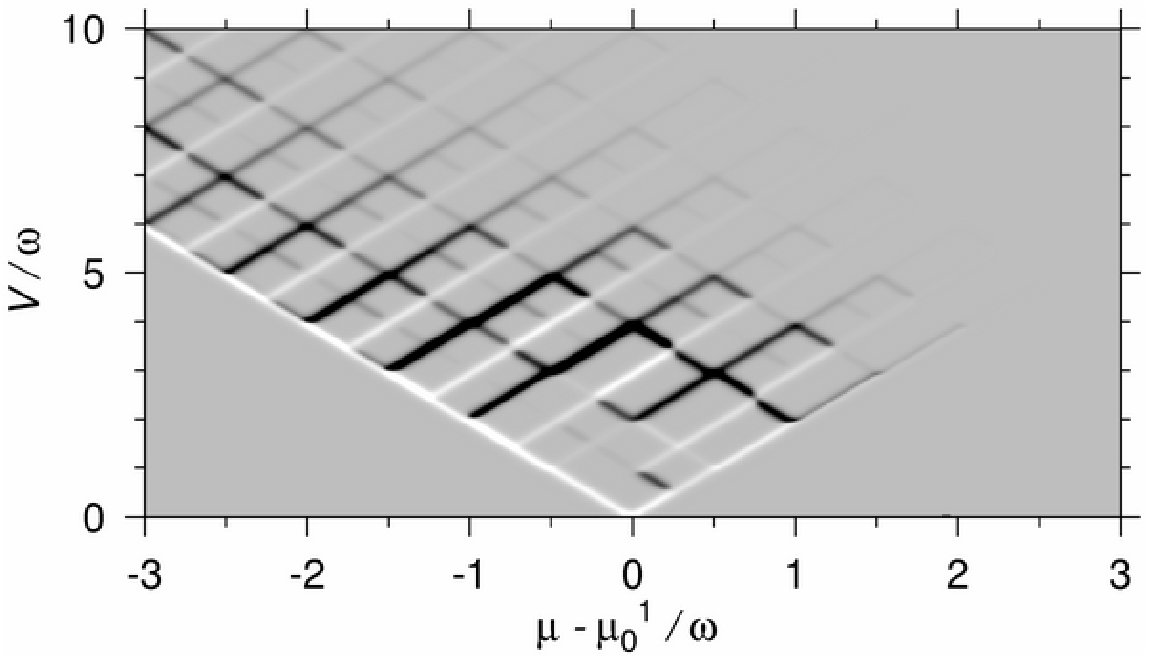}
   \caption{\label{fig:G_lLs}
     $dI/dV(\mu,V)$ for
     $\lambda_1=1.1$, $\lambda_2=5.0$ and $\Delta/\omega=0.5$.
   }
 \end{figure}
 \begin{figure}
   \includegraphics[scale=0.7]{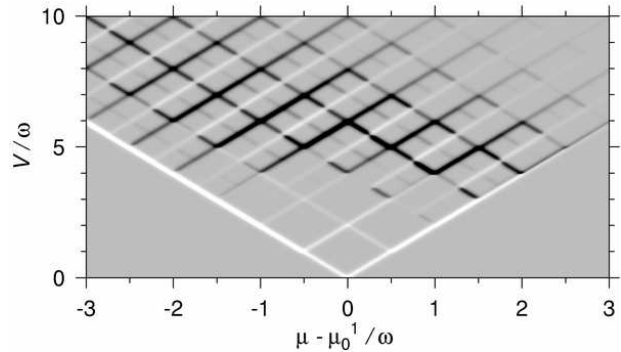}
   \caption{\label{fig:G_lLl}
     $dI/dV(\mu,V)$ for
     $\lambda_1=1.1$, $\lambda_2=5.0$ and $\Delta/\omega=2.5$.
   }
 \end{figure}
 Indeed, a glance at Figs.~\ref{fig:G_Lls}-\ref{fig:G_lLl} already shows that
 more NDC lines are visible.
 Also, there are more NDC lines with positive slope, which proves that
 the deviations from an equilibrium vibrational distribution are
 stronger.
 For $\lambda_1^2 \gg \lambda_2^2 > 1$ the NDC effects are strongest
 for the case $\Delta < \omega$ presented in Fig.~\ref{fig:G_Lls}.
 Compared with Fig.~\ref{fig:G_LSs} the isolated region
 defined by strong NDC lines is repeated a
 number of times to the left and it extends further to the right.
 Also, the current peak at the resonance line $\mu_L =\mu^2_0$ has shifted
 further to the left.
 The extended $I(V)$ plateaus have a width fixed by $\omega-\Delta$,
 independent of the gate voltage
 (compare with the weak NDC in Sec.~\ref{sec:lS} where the current steps and
 drops have opposite gate-voltage dependence).
 For the case $\Delta > \omega$ presented in Fig.~\ref{fig:G_Lll} the
 low bias structure disappears and the current peak along $\mu_L =
 \mu^2_0$ becomes the dominant feature.
 For $\lambda_2^2 \gg \lambda_1^2 > 1$ the NDC effects are strongest for
 the case $\Delta > \omega$ presented in Fig.~\ref{fig:G_lLl}.
 The two strong NDC lines in Fig.~\ref{fig:G_SLl}  have developed into a
 ``checkerboard'' pattern of such lines.
 These correspond to excitations of the weaker coupled state.
 In addition more resonances due to the strongly coupled state
 appear.
 For $\Delta < \omega$ the current is more suppressed
 at positive gate energies, Fig.~\ref{fig:G_lLs}.
\section{Summary and discussion
\label{sec:discuss}}

 We have calculated the non-linear current through a molecule with two
 non-degenerate electron-accepting orbitals coupled asymmetrically to
 an internal vibration in the limit of weak tunneling to the
 electrodes.
 We found that due to the interplay of Coulomb blockade and
 non-equilibrium vibration-assisted tunneling
 NDC effects become amplified and pervasive in comparison with a
 one-orbital model.
 The only resonances where we consistently find current steps
 correspond to an electron tunneling off the molecule starting from the
 charged state coupled strongly to the vibration.
 At all other resonance lines the $dI/dV$ may become negative depending on
 the electron-vibration couplings and applied voltages.
 A weak and strong NDC effect may be distinguished, which require the
 larger of the two electron-vibration couplings to be moderate and strong
 respectively.
\\
 The { weak NDC} effect is found at resonance lines where an electron
 can tunnel onto the molecule resulting in the charged state coupled
 {\em stronger} to the vibration.
 This effect only occurs when two (or more) orbitals are
 { competing} in the transport.
 The current steps and drops occur at bias positions with an {\em
   opposite} gate-voltage dependence and give rise to current {\em
   oscillations}.
 We proved that this type of NDC is robust against strong relaxation of the
 vibrational distribution on the molecule  due to a dissipative
  environment.
 Any NDC at other resonances conditions is a proof
 of a non-equilibrium vibrational distribution on the molecule.
 The current oscillations may even become amplified by strong
 relaxation depending on the applied voltages.
 \\
 { Strong NDC} effects appear at the first few resonance lines
 associated with the state {\em weakly} coupled to the vibration.
 This is a non-equilibrium effect which is typically weakened by
 relaxation processes.
 { Cascades} of single-electron tunneling processes
 involving the vibrational excitations of the weakly coupled state
 provide a {\em feedback} which rapidly populates the
 strongly coupled state.
 The latter thus acts as a {blocking} state which is almost fully
 occupied.
 The few, strong NDC lines correspond to the activation of the
 feedback mechanism and can have the {\em same} gate voltage dependence
 as the current steps,
 in contrast to the weak NDC above.
 In a one-orbital model the feedback mechanism is also active but
 produces only a weak effect due to the absence of a competing orbital.
 An anomalous {\em current peak} of width $\propto T$
 appears when
 the feedback mechanism becomes effective only sufficiently
 deep {\em inside} a resonance.
 The peak signals the crossover of the vibrational distribution
 from equilibrium to non-equilibrium.
\\
 Interestingly, the {blocking} state can be the vibrational ground
 state of either charged state whichever couples
 stronger to the vibration.
 When the electronic excited orbital couples most strongly
 the NDC signals a voltage-controlled {\em
   population inversion} between the charged states
 induced by the vibration-assisted tunneling.
 Also, new resonances appear associated with an electron {
   entering} the molecule and {\em absorbing} vibrational energy
 stored on it (despite the low temperature)
 where the current is {\em suppressed}.
\\
 The NDC effects are due to asymmetry of the  orbital
 energies and couplings to the internal vibration which are {\em intrinsic}
 properties of the molecule. One can thus tailor the electronic
 response of the device by molecular engineering.
 In contrast to other NDC effects, we do not require detailed
 assumptions of orbital- and/or electrode- specific
 electronic wave-function overlap with the
 electrodes~\cite{Hettler02,Hettler02nanobio,Hettler03,Thielmann04a}
 nor bias-voltage dependent coupling to the
 vibration~\cite{McCarthy03}.
\\
 For the interpretation of transport spectroscopy experiments on
 molecular devices an important result of our work is that multiple
 orbitals may be relevant for effects at voltages where only a single
 orbital would seem to matter.
 For instance, an excited orbital may already dominate the
 transport by cascades of single tunneling events at a low voltage
 where this state is not directly accessible from the neutral ground state,
 cf. Fig.~\ref{fig:G_SLl}.
 Similarly the charged ground state may completely dominate the transport even
 at a high bias voltage where a far ``better conducting'' excited
 orbital is already directly accessible
 cf. Fig.~\ref{fig:G_LSs}.
\\
 We have used a basic parameterization of the nuclear potential
 surface of the electronic states and some comments are appropriate.
 For one, the nuclear potential shape may be anharmonic in the
 coordinate $Q$ considered here.
 The resulting qualitative changes may be determined from the
 FC-factors for these potentials, when plotted in similar fashion as
 Fig.~\ref{fig:parabola}.
 The main results are not sensitive to the fine details of these factors
 but only to their large-scale dependence on the vibrational
 numbers due to the shift of the potentials,
 which can be established by quasi-classical considerations.
 Secondly, anharmonic terms in the nuclear potential may also couple the
 mode $Q$ to other internal modes which we have not considered here.
 When this coupling is strong for a large number of such other modes
 or a Fermi- (nonlinear-)resonance is involved,
 intramolecular vibrational energy redistribution may relax the vibration.
 Generally, this will become more important for large molecules.
 The effect of relaxation has been discussed. If however,
 only a few other modes couple strongly to $Q$, say one, an
 interesting two mode problem occurs.
 A treatment of the effects of such multi-mode
 dynamics~\cite{Koeppel84} on the tunneling transport, lies outside the
 scope of the present paper.
\begin{acknowledgements}
  We acknowledge stimulating discussions with H. Schoeller, K. Flensberg,
  W. Belzig, I. Sandalov, and M. Hettler.
  M. R. W. acknowledges the financial support provided through the
  European Community's Research Training Networks Program under contract
  HPRN-CT-2002-00302, Spintronics.
\end{acknowledgements}
\appendix
\section{Strong relaxation
\label{app:relax}}
 We consider the limit $\Gamma \ll \gamma \ll T$ where the vibrational
 excitations have completely relaxed before each tunneling event due to
 a dissipative environment. The tunneling rates $\Gamma^{ir}$ are not
 assumed to be symmetric.  With the factorization ansatz
 $P_q^i=P^i P_q$,
 $P_q=e^{-q\omega/T}(1-e^{-\omega/T})$
 we can reduce equations (\ref{eq:dotP}) to an effective
 electronic three-level problem with voltage dependent rates~(\ref{eq:rate_eq}).
 The stationary probabilities and the current can be explicitly given
 ($r=R,L$):
 \begin{eqnarray}
   \frac{P^i}{P^0} =          2\frac {W_{i \leftarrow 0}}
                                     {W_{0\leftarrow i}} ,
 P^0=  \left[ 1+2 \sum_{i}\frac{W_{i \leftarrow 0}}
                                     {W_{0\leftarrow i}}
       \right]^{-1},
 \\
   I_r=\sum_{i} \left(
   2W_{i\leftarrow 0}^{r} P^{0}
  - W_{0\leftarrow i}^{r} P^{i}
                    \right).
  \end{eqnarray}
 We can now find a simple explicit condition for the occurrence of
 current steps or drops with increasing bias voltage.
 Consider an increase of the positive bias $V\rightarrow V'$ such that
 one additional transition involving orbital~$i$ comes into the bias window
 through a resonance with electrode $r=L,R$, $\mu_r = \mu^i_k$ for some
 $k=0,\pm 1, \pm 2,\ldots$ (for $V<0$ interchange $L\leftrightarrow R$ below).
 For simplicity we consider the values of the current on the two
 subsequent plateaus: only two  transition rates are then changed,
$W_{i \leftarrow 0}  \rightarrow W'_{i \leftarrow 0}$ and
$W_{0 \leftarrow i}  \rightarrow W'_{0 \leftarrow i}$
 the changes being related by
 $ \delta W_{0 \leftarrow i}^{r} =-
   e^{ - k \beta \omega}
   \delta W_{i \leftarrow 0}^{r}$
 (cf. Eqs.~\ref{eq:rate}, \ref{eq:rate_eq}).
 The change in the  stationary current $\delta I_r= I'_r-I_r$
 may be calculated at either electrode $r=L,R$ from~(\ref{eq:Ieq})
 (since $I_L+I_R=I'_L+I'_R=0$):
 \begin{eqnarray}
\label{eq:dI/I}
   \frac{ \delta I_{r} }{ I_{r} }
   =2 \left(
     \frac{ W'_{i \leftarrow 0} }{ W'_{0 \leftarrow i} }
     -\frac{ W_{i \leftarrow 0} }{ W_{0 \leftarrow i} }
   \right)  P'^0
      \left(
      \frac{W^{\bar{r}}_{0 \leftarrow i}}{I^{r}}-1
     \right),
 \end{eqnarray}
 Here $\bar{r}=R,L$ denotes the electrode opposite to $r=L,R$.
 In order to have NDC at a resonance where $\mu_L$ becomes larger than
 $\mu^i_k$ we require
 $\delta I_{L} / I_{L} < 0$, which, using
 $r=L$ in Eq.~(\ref{eq:dI/I}),
 gives
 $W^{R}_{0 \leftarrow i} - I_{L} < 0$.
 The rate of escape through junction $R$ at voltage $V'$
 (including the new transition in increased bias window $V'$)
 must thus be smaller than the current $I_L$ at initial voltage $V$.
 It is readily seen that this condition cannot be fulfilled in the case 
 of only one orbital:
 $W^{R}_{0 \leftarrow 1} - I_{L} =
  W^{R}_{0 \leftarrow  1} + I_{R}
  =2 W^{R}_{1 \leftarrow 0}P^0
   + W^{R}_{0 \leftarrow  1}(1-P^1) > 0$.
 However, for two (or more) orbitals it is possible to satisfy this requirement.
 To see if NDC may occur at a resonance where $\mu_R$ drops below
 $\mu^i_k$ we use $r=R$ in Eq~(\ref{eq:dI/I}), leading to the
 requirement
 $W^{L}_{0 \leftarrow i}-I_R = W^{L}_{0 \leftarrow i}+I_L < 0$.
 which can not be satisfied for any applied voltages since $I_L>0$ for $V>0$.
 The current {\em must} increase at such resonances.
 Thus for fully equilibrated vibrations NDC can only occur at resonances
 related to the left electrode,
 $\mu_{L}=\mu-\mu^i_{k},k=0,\ldots$,
 where electrons can enter the molecule by an additional tunneling process
 $i_{q+k}\leftarrow 0_q$.
 These resonances correspond to lines with positive slope in the
 $(\mu,V)$ plane.
 Any NDC occurring at an other resonance is a proof of a
 non-equilibrium vibrational distribution.
 This proof can be trivially extended to $N$ orbitals correlated by
 Coulomb charging (maximally one extra electron). It also does not
 depend on the FC-factors involved, although the amplitude of the
 possible NDC may be small for a particular choice.
\section{Uncorrelated vibration-assisted tunneling
\label{app:uncorr}}
 We consider the special case where the renormalization of the
 interaction (due to the polaron effect) compensates the Coulomb
 repulsion effects, i.e. $v=v^{(0)}-2\omega \lambda_1\lambda_2=0$.
 Now we have to include the doubly charged state of the molecule
 $n_1=1,n_2=1$ (di-anion).
 We denote the diagonal density matrix elements by
 $P_{q}^{n_1n_2}$, where $q$ is the vibrational number and $n_1n_2$
 denotes an electronic state with occupations $n_i=0,1$ of orbital
 $i=1,2$.
 The occupations are coupled by the
 stationary master equations (cf.~\cite{Wegewijs99})
 \begin{eqnarray}
  &&
 -\sum_{i}  \left[
     n_i  \sum_{q'}   W_{0 q'\leftarrow i q}
 +(1-n_i) \sum_{q'} 2 W_{i q'\leftarrow 0 q}
 \right]                              P_{q}^{n_{1}n_{2}} \nonumber \\
 &&
 +\sum_{q'} \left[
      2 n_1   W_{1 q \leftarrow 0 q'} P_{q'}^{0n_{2}}
   +  2 n_2   W_{2 q \leftarrow 0 q'} P_{q'}^{n_{1}0}   \right. \nonumber \\
 &&
 \left.
   +(1-n_1) W_{0 q\leftarrow 1 q'} P_{q'}^{1n_{2}}
   +(1-n_2) W_{0 q\leftarrow 2 q'} P_{q'}^{n_{1}1}
 \right]=0
 \nonumber
 \end{eqnarray}
 together with $\sum_{n_1n_2} P^{n_1n_2}=1$.
 In the case where the vibration is assumed to be completely
 equilibrated, $\bar{P}^{n_{1}n_{2}}_q = \bar{P}^{n_{1}n_{2}} P_{q}$
 the kinetic equations can be decoupled into equations for the
 occupations of two uncorrelated ``channels'',
 $\bar{P}^i \equiv \sum_{n_1 n_2} \delta_{1n_{i}} P^{n_{1}n_{2}}$ with
 the rates (\ref{eq:rate_eq}):
 $  \bar{P}^{i} =0
   = 2W_{i\leftarrow 0} (1-\bar{P}^i)
     -W_{0\leftarrow i}    \bar{P}^i
 $.
 The current is then a sum of independent contributions of the
 individual orbitals:
 $I_r=\sum_{i} I_{ri}$ where
 $ I_{ri}= ( {   W^r_{0\leftarrow i} }^{-1}
           + { 2 W^r_{i\leftarrow 0} }^{-1} )^{-1}
 $.
 For a single orbital the current monotonically increases in the limit of strong
 relaxation (Appendix~\ref{app:relax}) and the same thus holds for two
 (or more) uncorrelated orbitals.

\end{document}